\documentclass[12pt]{article}
\usepackage{amsfonts}
\usepackage{amsmath,amssymb}
\usepackage{array}
\usepackage{color}
\usepackage{amsmath}
\usepackage{amsxtra}
\usepackage{amstext}
\usepackage{amssymb}
\usepackage{latexsym}
\newtheorem{thm}{Theorem}[section]
\newtheorem{prop}[thm]{Proposition}
\newtheorem{lem}[thm]{Lemma}
\newtheorem{remark}[thm]{Remark}
\newtheorem{conjecture}[thm]{Conjecture}

\newtheorem{corollary}[thm]{Corollary}
\numberwithin{equation}{section}
\def\Tr{{\rm Tr\,}}

\def\Det{{\rm Det}}
\def\R{\Bbb R}
\def\C{\Bbb C}
\def\Z{\Bbb Z}
\def\N{\Bbb N}

\def\supp{{\rm supp\,}}

\begin{document}

\phantom{.} {\qquad \hfill \textit{\textbf{{Draft: July 16, 2008 }}}}

\vskip 2.5cm

\begin{center}

{\Large{\bf {Mean-Field Interacting Boson Random Point Fields}}}

\smallskip

{\Large{\bf {in Weak Harmonic Traps}}}

\vskip 1cm

\setcounter{footnote}{0}
\renewcommand{\thefootnote}{\arabic{footnote}}

{\textbf{Hiroshi Tamura} \footnote{tamurah@kenroku.kanazawa-u.ac.jp}\\
       Graduate School of the Natural Science and Technology\\
           Kanazawa University,\\
          Kanazawa 920-1192, Japan

\bigskip

\textbf{Valentin A.Zagrebnov }\footnote{zagrebnov@cpt.univ-mrs.fr}\\
Universit\'e de la M\'editerran\'ee(Aix-Marseille II) and  \\
Centre de Physique Th\'eorique - UMR 6207 \\
CNRS-Luminy-Case 907, 13288 Marseille Cedex 9, France}

\vspace{1.5cm}

\end{center}

\begin{abstract}
A model of the mean-field interacting boson gas trapped by a weak harmonic potential is considered
by the \textit{boson random point fields} methods.
We prove that in the Weak Harmonic Trap (WHT) limit there are two phases distinguished by the
boson condensation and by a different behaviour of the local particle density.
For chemical potentials less than a certain critical value, the resulting Random Point
Field (RPF) coincides with the usual boson RPF, which corresponds to a non-interacting
(\textit{ideal}) boson gas.
For the chemical potentials greater than the critical value, the boson RPF
describes a divergent (local) density, which is due to \textit{localization} of the macroscopic number of
condensed particles.
Notice that it is this kind of transition that observed in experiments producing the Bose-Einstein
Condensation in traps.
\end{abstract}

\vspace{1.5cm}


\newpage
\tableofcontents

\vspace{1.5cm}

\section{Introduction and Main Results}
\subsection{{Weak Harmonic Traps} }
We consider the quantum statistical mechanical models of boson gas equipped with
a $\kappa$-parameterized family of one-particle Hamiltonians of harmonic oscillators:
\begin{equation}\label{Harm-Ham}
h_{\kappa} = \frac{1}{2}\sum_{j=1}^d \bigg( -\frac{\partial^2}{\partial x_j^2} + \frac{x_j^2}{\kappa^2}
-\frac{1}{\kappa}\bigg),
\end{equation}
which are self-adjoint operators in the Hilbert space $\mathfrak{H}:=L^2(\R^d)$.

In this setup a "thermodynamic limit" corresponds to $\kappa \rightarrow \infty$ {\rm{(}}i.e. the "opening"
of the trap {\rm{\cite{DGPS}}}{\rm{)}}, which we call the Weak Harmonic Trap (WHT) limit. Notice that the set
$C_{0}^{\infty}(\R^d)$ is a form-core of the operator (\ref{Harm-Ham}) and that this set is also a form-core for
the operator $(-\Delta)/2$. Here $\Delta$ is the standard Laplace operator in $\R^d$.
Then {\rm{(}}see e.g. {\rm{\cite{Ka}}}{\rm{)}} one obtains the strong resolvent convergence:
\begin{equation}\label{str-res-conv}
\lim_{\kappa \rightarrow \infty} h_{\kappa} = ( - \Delta)/2 \ .
\end{equation}
In spite of convergence (\ref{str-res-conv}),
there is a drastic difference between the properties of the infinite Ideal Boson Gas (IBG) systems
prepared via the WHT limit and via standard \textit{thermodynamic limit} (TDL) $\lim_{L\rightarrow\infty}\Lambda_L = \R^d$
with the one-particle operators ${t}_{L}:= (-\Delta/2)_L$ with a "non-sticky" (e.g. Dirichlet) boundary conditions
{\rm{\cite{VVZ}}}. Here $\Lambda_{L=1}\subset \R^d$ is a bounded open region of unit volume $|\Lambda_{L=1}| =1$
containing the origin $x=0$ whose boundary $\partial \Lambda_{L=1}$ is piecewise continuously differentiable and
\begin{equation}\label{Volume}
\Lambda_L := \{x\in \mathbb{R}^d | L^{-1} x \in \Lambda_{L=1}\} \ , \ L > 0  .
\end{equation}
In fact, it is known that the Bose-Einstein Condensation (BEC) occurs for dimensions $d>1$
in the IBG via
WHT limit $\kappa \rightarrow \infty$, while for dimensions $d>2$ in the IBG via TDL,
see (\ref{crit-dens-Ideal}) and {\rm{\cite{DGPS}, \cite{PeSm}}}.
Similarly, it is well-known that thermodynamic properties of the boson gases are very
sensible to different ways of taking the thermodynamic limit {\rm{\cite{LePu}}} or
to the different choices of the boundary conditions {\rm{\cite{VVZ}}}.

The purpose of the paper is to examine the position distribution of the mean field boson model
in WHT limit and to compair its behaviour to those of the mean field boson models in TDL
or ideal boson models in WHT limit.


Our method is based on the theory of Random Point Fields (RPFs) (see e.g. \cite{DV}).
The usual boson and the fermion RPFs  \cite{Ly, M75, M77} have been formulated in a unified way in terms of
the \textit{Fredholm determinant} together with other related RPFs, which are indexed by fractional numbers,
in \cite{ST03}.
They have been re-derived as theories which describe position distributions of the constituent particles
of quantum gases in the thermodynamic limit for canonical ensembles in \cite{1}.
It was shown that the Random Point Field (RPF) corresponding to fractional numbers \cite{ST03}, describes the
gases which consist of particles obeying the parastatistics \cite{3}.
The RPF describing a homogeneous Bose-Einstein condensation have been studied for the first time in \cite{2},
where the RPF is given by the convolution of usual boson RPF and another RPF.
The latter one seems to describe position distribution of the condensed part of the constituent bosons.
This RPF has been re-formulated using the Cox process.\cite{EK}

These theories of RPFs yield a precise information about the \textit{position} distribution of the constituent
quantum particles, although they are not suitable to characterize the quantum systems completely
(however see \cite{FF91, F91}).

\subsection{{Ideal and Mean-Field Boson Gases in the WHT} }
Recall that in the  \textit{grand-canonical} Gibbs ensemble the partition function of the IBG trapped by
harmonic potential (\ref{Harm-Ham}) is given by
\begin{equation}\label{Gr-Can-Part-Ideal}
\Xi_{0, \kappa}(\beta, \mu) := \sum_{n=0}^{\infty}e^{\beta \mu n }
\ \Tr_{\mathfrak{H}^{n}_{symm}}[\otimes^n G_{\kappa}(\beta)] \ .
\end{equation}
Here  $\mathfrak{H}^{n}_{symm}:= (\otimes^n L^2(\R^d))_{symm}$ is the $n$-fold \textit{symmetric}
Hilbert space tensor product of $\mathfrak{H}:= L^2(\R^d)$,
$G_{\kappa}(\beta) = e^{-\beta h_{\kappa}}$ the one particle Gibbs semigroup.
The zeroth term in (\ref{Gr-Can-Part-Ideal}) equals to 1 by definition.
We consider the case of positive inverse temperature $ \beta >0$ and of negative chemical
potential $\mu <0$.

The spectrum of the operator (\ref{Harm-Ham}) is discrete and has the form:
\begin{equation}\label{spec-WHT}
\mbox{Spec}(h_{\kappa}) =
\{\, \epsilon_{\kappa}(s):=|s|_1/\kappa \, | \, s =(s_1, \cdots, s_d) \in \Z_+ ^d \,\}
\end{equation}
where $|s|_1 := \sum_{j=1}^d s_j$ and $ \Z_{+}^1 $ is the set of all non-negative
integers. The normalized eigenfunctions of the states for $s \in \Z_{+}^d$ are given by
\begin{equation}\label{Hermite-d}
\phi_{s, \, \kappa} (x)= \frac{1}{\kappa^{d/4}}  \ \phi_{s} (x/\sqrt{\kappa})
          = \prod_{j=1}^d \ \frac{1}{\kappa^{1/4}}\phi_{s_j} (x_j/\sqrt{\kappa}) \ ,
\end{equation}
where for each component $j$, $\phi_{s_j}$ is related to the
\textit{Hermite} polynomials $H_{s_j}(z)$  by
\begin{equation}\label{Hermite}
\phi_{s_j} (x_j) =
(2^{s_j}\ s_{j}! \ \sqrt{\pi})^{- 1/2} H_{s_j}(x_{j})e^{-x_j^2/2} ,  \quad  j= 1,\ldots, d \ .
\end{equation}
The ground state is denoted in this paper by
\begin{equation}\label{Harmon-Ground}
     \Omega_{\kappa}(x)= \frac{1}{(\pi\kappa)^{d/4}}e^{-|x|^2/2\kappa}
     \equiv \phi_{s=0, \, \kappa} (x) \ ,
\end{equation}
where $ x=(x_1, \cdots, x_d) \in \R^d $, $ |x|^2 := \sum_{j=1}^dx_j^2$.

Integral kernel of  $G_{\kappa}(\beta) = e^{-\beta h_{\kappa}}$ has
the explicit form (the Mehler's formula for \textit{oscillator processes}):
\begin{equation}\label{Mehler}
G_{\kappa}(\beta; x,y)= \frac{\exp\{-(2\kappa)^{-1}{\tanh(\beta/2\kappa)}
(|x|^2 +|y|^2) -|x-y|^2/(2\kappa\sinh(\beta/\kappa))\}}
{\{\pi\kappa(1-e^{-2\beta/\kappa})\}^{d/2}} \ .
\end{equation}
Here the operator $G_{\kappa}(\beta)$ belongs to the
\textit{trace-class} $\mathfrak{C}_1 (L^2(\R^d)) $, with the trace-norm equals to
$\Tr {G_{\kappa}(\beta)} = 1/(1-e^{-\beta/\kappa})^d = O(\kappa^d)$ for large $\kappa$.
The largest eigenvalue of $G_{\kappa}(\beta)$ coincides with the operator norm
$ \left\| G_{\kappa}(\beta)\right\| = 1$. We write all the eigenvalues of operator $G_{\kappa}(\beta)$
in decreasing order:
\begin{equation*}
g_0^{(\kappa)}=1 > g_1^{(\kappa)} = e^{-\beta/\kappa}
\geqslant g_2^{(\kappa)} \geqslant \cdots.
\end{equation*}

The expectation value of \textit{total} number of particles is given by
\begin{equation}\label{total-number-Ideal}
         {N}_{\kappa}(\beta,\mu) = \frac{1}{\ \beta}
         \frac{\partial \ln \Xi_{0,\kappa} (\beta,\mu)}{\partial \mu}
            = \sum_{s \in \Z_+ ^d }
\frac{1}{e^{\beta(\epsilon_{\kappa}(s)- \mu )} - 1} \ .
\end{equation}
Since the value (\ref{total-number-Ideal}) diverges in the WHT limit
$\kappa\rightarrow\infty$ as $\kappa^{d}$ , one introduces the scaled quantity  \cite{DGPS}, \cite{PeSm}:
\begin{equation}\label{sum-density-Ideal}
\rho_{\kappa}(\beta,\mu):= \frac{1}{\kappa^{d}} \sum_{s \in \Z_+ ^d }
\frac{1}{e^{\beta(\epsilon_{\kappa}(s)- \mu )} - 1} \ ,
\end{equation}
which is a Darboux-Riemann sum for the integral
\begin{equation}\label{global-density-Ideal}
\rho(\beta,\mu)= \lim_{\kappa\rightarrow\infty} \rho_{\kappa}(\beta,\mu) =
\int_{[0, \infty)^d}  \ \frac{dp}{e^{\beta(|p|_1 - \mu )} - 1}=
\sum_{s=1}^{\infty} \frac{e^{\beta\mu s}}{(\beta s)^d}\ .
\end{equation}
Since $\kappa^d$ may be interpreted as {\it the effective ``volume"}
({\it cf.} Remark \ref{U-PHI}),
$\rho(\beta, \mu)$ and $\rho_{\kappa}(\beta, \mu)$ are regarded as the expectation value of
effective space-averaged density of the system, which has non-homogeneous space distribution.
One defines its critical value as usual:
\begin{equation}\label{crit-dens-Ideal}
             \rho_c (\beta):= \sup_{\mu<0}\rho(\beta,\mu) = \zeta(d)/\beta^d.
\end{equation}
Notice that (\ref{crit-dens-Ideal}) is bounded for $d>1$.
Therefore, if  $\rho > \rho_c (\beta)$,
the IBG in the WHT limit $\kappa\rightarrow\infty$ manifests a BEC in the scaled oscillator ground state
(\ref{Harmon-Ground}) with the expected  \textit{space-averaged} condensate density:
\begin{equation}\label{BEC-Ideal}
               \rho_{0}(\beta) := \rho - \rho_c (\beta) =
            \lim_{\kappa\rightarrow\infty} \ \frac{1}{\kappa^{d}} \frac{1}{e^{-\beta
         \overline{\mu}_{\kappa}(\beta,\rho)} - 1} \ .
\end{equation}
Moreover the expected local density $\rho_0(\beta)(x) $ can be defined such that
\begin{equation}\label{eld}
            \rho_{0}(\beta)(x) =  \lim_{\kappa\rightarrow\infty} \,
          \frac{1}{\kappa^{d}} \frac{\Omega_{\kappa}(x)^2}
      {e^{-\beta\overline{\mu}_{\kappa}(\beta,\rho)} - 1}
\end{equation}
holds. Here $\overline{\mu}_{\kappa}(\beta,\rho)$ is the unique root of the equation,
              $\rho = \rho_{\kappa}(\beta,\mu)$  \ {\it cf.} (\ref{sum-density-Ideal}).
Their limits satisfy
\begin{equation}\label{two-regimes}
        \lim_{\kappa\rightarrow\infty} \overline{\mu}_{\kappa}(\beta,\rho < \rho_c (\beta))< 0 \  \
      \textrm{and}\ \ \lim_{\kappa\rightarrow\infty}
     \overline{\mu}_{\kappa}(\beta,\rho \geqslant \rho_c (\beta))= 0 ,
\end{equation}
especially
\begin{equation}\label{asympt}
\overline{\mu}_{\kappa}(\beta,\rho > \rho_c (\beta))= - \frac{1}{\beta (\rho - \rho_{c}(\beta)) \kappa^{d}}
+ o (\kappa^{-d}) \ .
\end{equation}
The integrated density of states $\mathcal{N}_{\kappa}(E)$ for the operator  $h_{\kappa}$ is
given by
\begin{equation*}
\mathcal{N}_{\kappa}(E) = \frac{1}{\kappa^d}\sum_{s \in \Z_+ ^d }
\theta (E - |s|_1 / \kappa).
\end{equation*}
Then from its Laplace transform
\begin{equation*}
        \int_0 ^{\infty} e^{- t E} d\mathcal{N}_{\kappa}(E)
       = [\kappa (1 - \exp (-t/\kappa))]^{-d},
\end{equation*}
we obtain the $\kappa\rightarrow\infty$ limit
\begin{equation}\label{HT-state-dens}
d\mathcal{N}(E)= \frac{E^{d-1}}{\Gamma(d)} \, d E \ .
\end{equation}
In terms of these density of states, (\ref{sum-density-Ideal}) and
(\ref{global-density-Ideal}) are written as
\[
      \rho_{\kappa}(\beta, \mu) = \int_0 ^{\infty} \frac{ d\mathcal{N}_{\kappa}(E)}
            {e^{\beta (E-\mu)}-1}, \quad
        \rho(\beta, \mu) = \int_0 ^{\infty} \frac{ d\mathcal{N}(E)}{e^{\beta (E-\mu)}-1}.
\]

It is instructive to compare these results with properties of the IBG ``prepared" via standard thermodynamic limit
$L\rightarrow\infty$ (\ref{Volume}) for e.g. Dirichlet boundary conditions. It is well-known \cite{ZB} that in this case
the expected boson density is
\begin{equation}\label{loc-density-Stand}
\rho_{\Lambda_L}(\beta,\mu)= \frac{1}{|\Lambda_L|} \sum_{j \in \Z_+ }
\frac{1}{e^{\beta(\varepsilon_{L}(j)- \mu )} - 1} \ .
\end{equation}
Here $\{\varepsilon_{L}(j)\}_{j\geqslant 0}$ is the spectrum of the one-particle
operator $t_L$  and $\mu < \min_{j\geqslant 0}\varepsilon_{L}(j) \to  0 $ as $ L\to \infty$.
It also has the expression
\begin{equation}\label{total-dens-IBG}
               \rho_{\Lambda_L}(\beta,\mu) =
    \int_{0}^{\infty} \frac{d \widetilde{\mathcal{N}}_{L}(E)}{e^{\beta(E - \mu )} - 1}
\end{equation}
in terms of the \textit{integrated density of states} $\{\widetilde{\mathcal{N}}_{L}(E)\}_{L>0}$.
The thermodynamic limit $\widetilde{\mathcal{N}}(E)= \lim_{L\to \infty}\widetilde{\mathcal{N}}_{L}(E)$
is independent of "non-sticky" boundary conditions \cite{RSIV} and given by
\begin{equation}\label{stand-state-dens}
 d\widetilde{\mathcal{N}}(E) =  \frac{E^{(d-2)/2}}{(2\pi)^{d/2} \Gamma (d/2)} \, dE  \ .
\end{equation}
Then we get the limit of the expected density
\begin{equation}\label{total-dens-IBG-lim}
          \widetilde{\rho}(\beta,\mu)= \lim_{L\rightarrow\infty}\rho_{\Lambda_L}(\beta,\mu)=
         \int_{0}^{\infty} \frac{d \widetilde{\mathcal{N}}(E)}{e^{\beta(E - \mu )} - 1}  \ .
\end{equation}
The critical particle density for the IBG is
\begin{equation}\label{crit-dens-stand}
\widetilde{\rho}_c (\beta):= \sup_{\mu<0}\widetilde{\rho}(\beta,\mu)
= \int_0^{\infty}\frac{d\widetilde{\mathcal{N}}(E)}{e^{\beta E} -1}
= \zeta(d/2)/ (2\pi \beta)^{d/2} \ .
\end{equation}
Note the difference between $\widetilde{\rho}_c (\beta)$ and ${\rho}_c (\beta)$.
In particular (\ref{crit-dens-stand}) is bounded only when $d>2$.
Thus thermodynamic properties of the IBG in the standard TDL
$\Lambda_L \rightarrow {\mathbb{R}}^d$ and the WHT limit $\kappa \rightarrow \infty$ are different
in spite of delusive impression that they have to produce identical systems.

\smallskip

Now we consider the  mean-field interacting bosons trapped in the harmonic potential (\ref{Harm-Ham}).
Its {\it grand-canonical partition function} is given by
\begin{equation}\label{Gr-Can-Part}
\Xi_{\lambda, \kappa}(\beta, \mu) := \sum_{n=0}^{\infty}e^{\beta(\mu n - \lambda n^2/2\kappa^d)}
\ \Tr_{\mathfrak{H}^{n}_{symm}}[\otimes^n G_{\kappa}(\beta)] \ .
\end{equation}
We consider the case of $\beta>0, \lambda >0$ and arbitrary $\mu \in \R$.
Hereafter, we suppress the symbol $\lambda$ from the left-hand side of (\ref{Gr-Can-Part}),
since we fix $\lambda >0$ in the rest of the paper.

\begin{remark}\label{U-PHI}
The scaling with the "volume" $\kappa^d$ (\textit{imposed} by the WHT) is a
\textit{conventional} way to consider the Bose-Einstein condensation in traps, see {\rm{\cite{DGPS}}},
{\rm{\cite{LSSY}}}, {\rm \cite{PeSm}}.
Our definition of the MF interaction in WHT applies a  space-average
over the "volume" $\kappa^d$, which plays the same r\^{o}le as $|\Lambda_{L}|$ in the standard
mean field model where the interaction has the form $\lambda n^2/2|\Lambda_{L}|$, see e.g. {\rm{\cite{ZB}}}.

Notice that $\lambda >0$ corresponds to \textit{repulsive} mean-field (MF) particle interaction,
whereas $\lambda = 0$ is the case of the IBG (\ref{Gr-Can-Part-Ideal}).

In the present paper we consider in (\ref{Gr-Can-Part}) only the "square" mean-field repulsive interaction
$U_2 := \lambda n^2/2\kappa^{d}$ .
Although application of the Large Deviation technique makes it possible to consider also the case of the
general MF interaction $U_\Phi:= \kappa^{d} \Phi(n/\kappa^{d})$, where $\Phi: \mathbb{R} \to \mathbb{R}$
is a piece-wise differentiable continuous function bounded from below, see \rm{\cite{TZ}.}
\end{remark}

To study the \textit{non-homogeneous} condensation and the \textit{space} distribution of the constituent bosons
in the system (\ref{Gr-Can-Part}) we use the RPF {$\nu_{\kappa, \beta}$}, i.e., the probability
measure on the space of locally finite point measures with generating functional:
\begin{eqnarray}\label{EK1}
E_{\kappa, \beta, \mu}(f):&=& \mathbb{ E}_{\kappa, \beta, \mu}\big[e^{-\langle f, \xi\rangle}\big] \\
&=& \frac{1}{\Xi_{\kappa}(\beta,\mu)} \sum_{n=0}^{\infty}
e^{\beta\mu n-\beta\lambda n^2/2\kappa^d} \Tr_{\mathfrak{H}^{n}_{symm}}
[\otimes^n(G_{\kappa}(\beta)e^{-f})] \ , \nonumber
\end{eqnarray}
where $ f \in C_0(\R^d), \; f \geqslant 0$.
Here $\mathbb{ E}_{\kappa, \beta, \mu}\big[ \, \cdot \, \big]$
stands for expectation with respect to $\nu_{\kappa, \beta, \mu}(d \xi)$,
and $\xi $ denotes the integral variable which represents locally finite point measure,
see \cite{DV,1,2,3}. The measure $\nu_{\kappa, \beta, \mu}$ describes a finite RPF whose \textit{Janossy measure}
can be given explicitly, see Remark \ref{Jano}.

In the present paper we study the properties of the MF {interacting} boson RPF $\nu_{\kappa,\beta,\mu}$
in the WHT limit $\kappa \to \infty$ by analyzing the generating functional (\ref{EK1}).
To this end we first define the MF \textit{critical} chemical potential
\begin{equation}\label{crit-chem-MF-WHT}
         \mu_{\lambda, c} (\beta):= \lambda \int_{[0,\infty)^d}\frac{dp}{e^{\beta|p|_1}-1}
         = \lambda \int_0 ^{\infty} \frac{ d\mathcal{N}(E)}{e^{\beta (E-\mu)}-1}
         =\frac{\lambda\zeta(d)}{\beta^d}.
\end{equation}
This critical parameter is similar to the critical chemical potential
$\widetilde{\mu}_c (\beta):= \lambda \, \widetilde{\rho}_c (\beta)$  for the
standard {homogeneous} MF boson gas via TDL, see e.g. \cite{ZB} and (\ref{crit-dens-stand}).

\subsection{{Main Results} }
Now we can formulate our main theorem:
\begin{thm}\label{thmA}
{\rm{(i)}} Let $\mu < \mu_{\lambda, c}(\beta)$ {\rm{(}}normal phase{\rm{)}}. Then the RPF
{$\nu_{\kappa,\beta,\mu}$} defined by (\ref{EK1}), converges weakly in the
WHT limit $\kappa \to \infty$ to the boson RPF {$ \nu_{\beta,r_*} $}
corresponding to the generating functional:
\begin{eqnarray}\label{norm}
{E}_{ \beta, r_*}(f): &=& \mathbb{E}_{ \beta, r_*} \big[e^{-\langle f, \xi\rangle}\big] \\
&=&\Det \big[1+\sqrt{1-e^{-f}}\ r_* \, G(\beta)(1-r_* \, G(\beta))^{-1}\sqrt{1-e^{-f}} \ \big]^{-1} \ .
\nonumber
\end{eqnarray}
Here $\mathbb{ E}_{ \beta, r_*}\big[ \, \cdot \, \big]$ denotes expectation with respect to
the measure $\nu_{\beta, r_*}$, $\Det$ stands for the {Fredholm determinant},
$G(\beta) := e^{\beta\Delta/2}$ is the {heat semigroup}
on $ \mathfrak{H}$ and $r_* = r_* (\beta,\mu,\lambda)\in (0,1)$ is a unique solution of the equation :
\begin{equation}\label{r-noBEC}
\beta\mu = \log r + {\lambda \beta} \int_{0}^{\infty}\frac{d\mathcal{N}(E)}{r^{-1} e^{\beta\, E}-1} \ .
\end{equation}
{\rm{(ii)}} For $\mu > \mu_{\lambda, c}(\beta)$ {\rm{(}}condensed phase{\rm{)}},
the  generating functional (\ref{EK1}) has the following asymptotics:
\begin{equation}\label{bec}
\lim_{\kappa\to\infty}\frac{1}{\kappa^{d/2}}
\log {E}_{\kappa, \beta, \mu}(f)
= - \ \frac{\mu- \mu_{\lambda, c}(\beta)}{{{\pi^{d/2}}}\lambda}
\ (\sqrt{1-e^{-f}}, (1+K_f)^{-1}\sqrt{1-e^{-f}}),
\end{equation}
where $K_f := \left(G(\beta)^{1/2}(1-G(\beta))^{-1/2}\sqrt{1-e^{-f}}\right)^*
 \left(G(\beta)^{1/2}(1-G(\beta))^{-1/2}\sqrt{1-e^{-f}}\right)$
is a positive trace-class operator on $ \mathfrak{H}= L^2(\R^d)$ for $d > 2$.
\end{thm}
\begin{remark}\label{dimension}
For dimensions $d > 1$ the integral (\ref{crit-chem-MF-WHT}) is {finite}.
This gives an idea that the BEC is possible
for the case $\mu > \mu_{\lambda, c}(\beta)$ and dimensions $d>1$.
However, in the present paper we assume $d>2$ to be able to prove our main theorem. In fact for $d=2$, the
operator $K_f$ is not a trace-class operator and the Fredholm determinant is not well-defined, see {\rm{\cite{2}}}.
\end{remark}
\begin{remark}
Because of a technical difficulty, actually we do not have results for the critical case
$\mu=\mu_{\lambda, c}(\beta)$.
\end{remark}
\begin{remark} \label{heat-kernal} The heat semigroup $G(\beta)$ is appeared here since
its kernel
\begin{equation*}
G(\beta; x,y) =(2\pi\beta)^{-d/2}e^{-|x-y|^2/2\beta}
\end{equation*}
is the point-wise limit of the Mehler kernel (\ref{Mehler}), as $\kappa \rightarrow \infty$.
Therefore the generating functional of the resulting RPF (\ref{norm}) in {\rm{(i)}} has exactly the same
form as that for the standard homogeneous IBG in the non-condensed phase, see eq.(2.13) in
{\rm{\cite{1}}}, see also {\rm{\cite{ST03}}}.
That is, the position distribution of the model in the non-condensed phase is coincides
with that of the standard IBG.
However, it is the integrated density of states $ \mathcal{N}(E) $ that is appeared in (\ref{r-noBEC}),
instead of $\widetilde{\mathcal{N}}(E)$.
It implies that the dependence of $r_*$ as a function of parameters, especially $\beta$,
is different from standard IBG or the standard MF-model.
\end{remark}
\begin{remark}\label{condensedRPF}
The RPF describing BEC for homogeneous IBG is given by the convolution of two RPFs {\rm{\cite{2}}}.
One convolution component is the usual boson RPF, while the other component seems to describe
the position distribution of the condensed part of the constituent bosons.
The behaviour of the generating functional in case (ii) shows that the latter component
overwhelms the former in the present model.
It is to be noted here that the latter can not be explained by the particles
in the ground state alone, it contains the effect of the interference between
``the condensed part" and ``the normal part" {\rm{\cite{EK}}}, although the intensity of the RPF is
proportional to the square of the ground state wave function, as (\ref{eld}).
\end{remark}

The sharp contrast between two regimes {\rm{(i)}} and {\rm{(ii)}} in Theorem \ref{thmA}
may be seen by the expectation values.

\begin{corollary} \label{comm-results-space}
For the case {\rm (i)} $\mu < \mu_{\lambda, c}(\beta)$ {\rm{(}}normal phase {\rm{)}}
\begin{eqnarray}\label{dens1}
\mathbb{E}_{ \beta, r_*}\big[\langle f, \xi\rangle \big] =
\Tr [f \ r_*G(\beta)(1-r_*G(\beta))^{-1}] = \rho_{r_*} \ \int_{{\mathbb{R}}^d} dx  \ f(x) \nonumber
\end{eqnarray}
holds, where $\rho_{r_*}$ is given by
\begin{eqnarray*}
\rho_{r_*}= r_*G(\beta)(1-r_*G(\beta))^{-1}(x,x) = \sum_{n=1}^{\infty}r_*^n/(2\pi\beta n)^{d/2}  .
\end{eqnarray*}
For the case {\rm (ii)} $\mu > \mu_{\lambda, c}(\beta)$ {\rm{(}}condensed phase{\rm{)}},
\begin{equation}\label{dens2}
           \liminf_{\kappa\to \infty}
  \frac{\mathbb{ E}_{\kappa, \beta, \mu, \lambda}\big[\langle f, \xi\rangle \big]}{\kappa^{d/2}}
 \geqslant \frac{\mu -\mu_{\lambda, c}(\beta)}{\pi^{d/2}\lambda} \int_{{\mathbb{R}}^d} dx \ f(x) \
\end{equation}
holds, where $ f \in C_0(\R^d)$.
\end{corollary}

The \textit{weak limits} of the RPFs concerns the limit of  the {\it local} position
distribution of particles.
In this sense, the results of Theorem \ref{thmA} and Corollary \ref{comm-results-space} in
regime (i) may be interpreted as follows: in the WHT limit the position distribution of
the MF interacting bosons in neighbourhoods of the origin of coordinates
(i.e. the bottom of the WHT potential) is close to that of
a free IBG corresponding to the unconventional parameter (\ref{r-noBEC}).
The information about the particle position distribution in domains distant from
the bottom of the WHT are missing in the limit $\nu_{\beta, r_*}$.
In order to take this ``tail"  particles into account we use the standard definition of the
grand-canonical total number of particles for our model :
\begin{eqnarray}\label{s-a-glob-dens}
{\rho}_{\kappa,\lambda}^{(tot)}(\beta,\mu)&:=& \frac{1}{\kappa^d \, \beta}
   \frac{ \partial \ln \Xi_{\kappa} (\beta,\mu)}{\partial \mu}\\
&=&
\frac{1}{\kappa^d \, \Xi_{\kappa, \lambda}(\beta, \mu)}
\sum_{n=0}^{\infty} n \, e^{\beta(\mu n - \lambda n^2/2\kappa^d)} \
\Tr_{\mathfrak{H}^{n}_{symm}}[\otimes^n G_{\kappa}(\beta)] \ . \nonumber
\end{eqnarray}
Since $\kappa^d$ is interpreted as the effective volume of the model,
$\rho^{(tot)}_{\kappa, \lambda}(\beta, \mu)$ represents the effective total \textit{space-averaged} density
of the non-homogeneous system (\ref{Gr-Can-Part}).

\begin{thm}\label{thmB}
The WHT limit of {\rm{(}}\ref{s-a-glob-dens}{\rm{)}}
\begin{equation}\label{Blim}
\rho^{(tot)}_{\lambda}(\beta, \mu) = \lim_{\kappa \to \infty} \rho^{(tot)}_{\kappa, \lambda}(\beta, \mu)
= \lim_{\kappa \to\infty}\kappa^{-d}\Tr[r_*G_{\kappa}(1-r_*G_{\kappa})^{-1}]
\end{equation}
exists and satisfies {\rm{:}}

\noindent {\rm{(i)}} {\rm{(}}$\mu \leq \mu_{\lambda, c}(\beta)${\rm{)}}
\begin{equation}\label{eq-tot-dens-noBEC}
    \rho^{(tot)}_{\lambda}(\beta, \mu) = \int_{[0, \infty)^d}
            \frac{d\mathcal{N}(E)}{r_*^{-1}e^{\beta E}-1} \quad  \mbox{ and } \quad
        \beta\mu = \log r_* + \lambda\beta\rho^{(tot)}_{\lambda}(\beta, \mu) \ ;
\end{equation}

\noindent {\rm{(ii)}} {\rm{(}}$\mu > \mu_{\lambda, c}(\beta)${\rm{)}}
\begin{equation}
\rho^{(tot)}_{\lambda}(\beta, \mu)= \mu/\lambda \ .
\end{equation}

It also holds that
\begin{equation}
\rho^{(tot)}_{c}(\beta):= \lim_{\mu \to\mu_c(\beta)}\rho^{(tot)}_{\lambda}(\beta, \mu) = \zeta(d)/\beta^d \ .
\end{equation}
\end{thm}
\begin{remark}\label{density-density}
The readers should not to confuse two ``densities" {\rm{:}} $\rho_{r_*}$ in Corollary
\ref{comm-results-space} and $\rho^{(tot)}_{\lambda}(\beta, \mu)$ defined above.
The $\rho_{r_*}$ can be interpreted as the limit of the ``local" density around the origin of
coordinate of non-homogeneous RPF $\nu_{\kappa, \beta, \mu}$, on the other hand the
$\rho^{(tot)}_{\lambda}(\beta, \mu)$ retains the information about the expectation  of the
total number of particles with respect to  $\nu_{\kappa, \beta, \mu}$ through the WHT limit.

Note that $ \rho^{(tot)}_{c}(\beta)$ coincides with (\ref{crit-dens-Ideal}) of IBG in the WHT.
\end{remark}
\begin{remark} \label{comm-results}
Qualitatively different behaviour of the space distributions of bosons described
in Theorem \ref{thmA} can be understand heuristically with the help of Theorem \ref{thmB}
in the following way. Consider the WHT limit $\kappa \rightarrow \infty$:

In case {\rm{(i)}} the bosons are distributed almost uniformly in the region of radius $\kappa$
according to the kernel (\ref{Mehler}).

On the other hand, in case {\rm{(ii)}} {\rm{(}}condensed phase{\rm{)}} the condensed part of
particles $\kappa^d(\rho^{(tot)}_{\lambda}(\beta,\mu) - \rho^{(tot)}_{\lambda,c} (\beta)) =
\kappa^d(\mu - \mu_{\lambda, c}(\beta))/\lambda$
is localized in the region of radius $O(\kappa^{1/2})$ according to the
profile of the square of the ground state wave function $\Omega_\kappa$.
Whereas the particles outside of the condensate essentially spread out over the region  of radius $\kappa$.
\end{remark}

The paper is organized as follows.
Preliminary estimates and results concerning the WHT limit for the \textit{mean-field} interacting boson gas
($\lambda > 0$) are collected in Section 2.
Section 3 and 4 are dedicated to the proof of Theorem \ref{thmA} and \ref{thmB}, respectively.
We reserved Section 5 for summary and conjectures.

\section{Preliminary Arguments and Estimates }

In this section, we write the expectation (\ref{EK1}) as the ratio
$\tilde\Xi_{\kappa}(\beta, \mu)/\Xi_{\kappa}(\beta, \mu)$.
The representations of $\tilde \Xi_{\kappa}(\beta, \mu)$ and $\Xi_{\kappa}(\beta, \mu)$ are
given in the form of integration of Fredholm determinants.
We also give the miscellaneous estimates needed for the evaluation of these integrals.
\subsection{ $\Xi_{\kappa}( \beta, \mu)$ and $\tilde\Xi_{\kappa}(\beta, \mu)$}
In terms of the projection operator on $\mathfrak{H}^{n}=\otimes^nL^2(\R^d)$
onto its subspace $\mathfrak{H}^{n}_{symm}$, the grand-canonical partition function can
be written as
\[
   \Xi_{\kappa}( \beta, \mu) = \sum_{n=0}^{\infty}e^{\beta\mu n-\beta\lambda n^2/2\kappa^d}
     \frac{1}{n!}\sum_{\sigma\in\mathcal{S}_n}
       \Tr_{\otimes^nL^2(\R^d)}\big[\big(\otimes^nG_{\kappa}(\beta)\big)U(\sigma)\big],
\]
where {the second sum is taken over the symmetric} group $\mathcal{S}_n$ and
\[
     U(\sigma) \varphi_1\otimes \cdots \otimes \varphi_n
       = \varphi_{\sigma^{-1}(1)}\otimes \cdots \otimes \varphi_{\sigma^{-1}(n)}
              \qquad \mbox{ for } \; \sigma \in \mathcal{S}_n, \;
                \varphi_1, \cdots, \varphi_n \in L^2(\R^d).
\]
Hence we have
\begin{eqnarray*}
      \Xi_{\kappa}( \beta, \mu) &=& \sum_{n=0}^{\infty}\frac{e^{\beta\mu n-\beta\lambda n^2/2\kappa^d}}{n!}
         \sum_{\sigma\in\mathcal{S}_n}\int_{(\R^d)^n}\big(\prod_{j=1}^n G_{\kappa}(\beta)
      (x_j, x_{\sigma^{-1}(j)})\big)dx_1\cdots dx_n \\
     &=& \sum_{n=0}^{\infty}\frac{e^{\beta\mu n-\beta\lambda n^2/2\kappa^d}}{n!}
       \int_{(\R^d)^n}{\rm Per}\,\big\{G_{\kappa}(\beta)
      (x_i, x_j)\big\}_{1\leqslant i,j\leqslant n}dx_1\cdots dx_n \ ,
\end{eqnarray*}
here {Per} stands for the permanent of the matrix $\big\{G_{\kappa}(\beta)
      (x_i, x_j)\big\}_{1\leqslant i,j\leqslant n}$ .
{\begin{remark}
The point field $\nu_{\kappa,\beta,\mu}$ of (\ref{EK1}) can also be defined in terms of Janossy
measures or exclusion probability {\rm{\cite{DV}}}.
This means that $\nu_{\kappa,\beta,\mu}$ is a finite point field, which assigns the
probability
\[
{\rm{Pr}}\{dX_n\}:= \frac{e^{\beta\mu n-\beta\lambda n^2/2\kappa^d}}{n! \ \Xi_{\kappa}(\beta,\mu)} \
      {\rm Per}\,\big\{G_{\kappa}(\beta)
      (x_i, x_j)\big\}_{1\leqslant i,j\leqslant n}dx_1\cdots dx_n
\]
to the event $\{dX_n\}$: there are exactly $n$ points, one in each infinitesimal region \\
$\prod_{i=1}^d(x_j^{(i)}, x_j^{(i)}+dx_j^{(i)}), \quad
\big( x_j=(x_j^{(1)}, \cdots, x_j^{(d)}), \; j=1, \cdots, n \big)$.
\label{Jano}
\end{remark}}
As in \cite{1,2}, we use the generalized \textit{Vere-Jones' formula} \cite{ST03,V} in the form
\[
     \frac{1}{n!}\int
                  {Per}\,\{J(x_{i}, x_{j})\}_{1\leqslant i,j \leqslant n}
                      \ dx_{1}\cdots dx_{n}
      = \oint_{S_r(0)}\frac{dz}{2\pi iz^{n+1}\Det(1- z J)},
\]
where $r>0$ satisfies $||r J||<1$.
$S_r(\zeta)$ denotes the integration contour defined
by the map  $ \theta \mapsto \zeta + r\exp(i\theta) $,
where $\theta$ ranges from $-\pi$ to $\pi$, $ r>0 $ and $ \zeta \in \C$.
Then we obtain
\[
     \Xi_{\kappa}( \beta, \mu) = \sum_{n=0}^{\infty}e^{\beta\mu n-\beta\lambda n^2/2\kappa^d}
    \oint _{S_r(0)}\frac{dz}{2\pi iz^{n+1}\Det(1- z G_{\kappa}(\beta))},
\]
where $ r \in(0, ||G_{\kappa}(\beta)||^{-1}) = (0,1) $.
Note that the zeroth term is 1 in this expression.

Let us substitute
\begin{equation}\label{Gauss-linear}
     e^{-\beta\lambda n^2/2\kappa^d}=\sqrt{\frac{\beta\lambda}{2\pi\kappa^d}}
    \int_{\R}dx\,\exp\Big(-\frac{\beta\lambda}{2\kappa^d}
         ((x+is)^2 - 2in(x+is))\Big).
\end{equation}
If $ s>0$ satisfies
\[
      e^{\beta\mu -\beta\lambda s/\kappa^d}< r,
\]
we can take the summation over $n$ together with the complex integration
and a scaling of $x$ to get
\[
     \Xi_{\kappa}(\beta, \mu) = \sqrt{\frac{\beta\lambda}{2\pi\kappa^d}} e^{\beta\lambda s^2/2\kappa^d}
    \int_{\R}dx\oint\frac{dz}{2\pi i}\frac{\,e^{-\beta\lambda (x^2 + 2isx)
         /2\kappa^d}}
  {\big(z-e^{\beta\mu+\beta\lambda(ix-s)/\kappa^d}\big)\Det[1-zG_{\kappa}(\beta)]}
\]
\begin{equation}
       = \sqrt{\frac{\kappa^d}{2\pi\beta\lambda}} e^{\beta\lambda s^2/2\kappa^d}
    \int_{\R}dx\,\frac{e^{-isx-\kappa^dx^2/2\beta\lambda}}
  {\Det[1- e^{\beta\mu +ix - \beta\lambda s/\kappa^d}G_{\kappa}(\beta)]}.
\label{pf1}
\end{equation}
Note that after $z$-integration, $r$ disappears and (\ref{pf1}) is valid for any $s$ satisfying
$   \exp(\beta\mu-\beta\lambda s/\kappa^d) \in (0, || G_{\kappa}(\beta)||^{-1} )   $.
We will estimate the integral in the spirit of saddle point method.
Here, we extract the main part from the integral.
Let $s = s_{\kappa}, r = r_{\kappa}$ be the unique solution of the system:
\begin{equation}
\begin{cases}
            r = \exp\big(\beta\mu - \beta\lambda s/\kappa^d )  \\
            s = \Tr [rG_{\kappa}(\beta)(1-rG_{\kappa}(\beta))^{-1}].
\end{cases}
\label{rs}
\end{equation}
Obviously, the condition $ r_{\kappa} \in (0, ||G_{\kappa}(\beta)||^{-1})$ is fulfilled.
Hence, we can substitute in (\ref{pf1}) $s$ by $s_{\kappa}$.
Using the product property of the Fredholm determinant, we get for denominator of (\ref{EK1}) the
representation:
\begin{equation}
     \Xi_{\kappa}(\beta, \mu) = \sqrt{\frac{\kappa^d}{2\pi\beta\lambda}}
        \frac{e^{\beta\lambda s_{\kappa}^2/2\kappa^d}}{\Det[1-r_{\kappa}G_{\kappa}(\beta)]}
    \int_{\R}dx\,\frac{e^{-is_{\kappa}x-\kappa^dx^2/2\beta\lambda}}
  {\Det[1- (e^{ix}-1)r_{\kappa}G_{\kappa}(\beta)(1-r_{\kappa}G_{\kappa}(\beta))^{-1}]}
\label{pf2}
\end{equation}
For the numerator of (\ref{EK1}), we introduce  bounded symmetric operators
\begin{equation}\label{G-f}
\tilde G_{\kappa}(\beta)(f): = G_{\kappa}(\beta)^{1/2} e^{-f} G_{\kappa}(\beta)^{1/2} \ ,
\end{equation}
indexed by function $ f \in C_0(\R^d), \; f \geqslant 0$, which we skip
below for simplicity. Then for generating functional (\ref{EK1}) one gets the form:
${E}_{\kappa, \beta, \mu}(f)= \tilde\Xi_{\kappa}(\beta, \mu)(f)/\Xi_{\kappa}(\beta, \mu) $, where
\[
      \tilde\Xi_{\kappa}(\beta, \mu)(f)= \tilde\Xi_{\kappa}(\beta, \mu)= \sum_{n=0}^{\infty}
      e^{\beta\mu n-\beta\lambda n^2/2\kappa^d}
     \Tr_{\mathfrak{H}^{n}_{symm}}[\otimes^n\tilde G_{\kappa}(\beta)]
\]
\begin{equation}\label{tpf1}
    = \sqrt{\frac{\kappa^d}{2\pi\beta\lambda}}
        \frac{e^{\beta\lambda \tilde s_{\kappa}^2/2\kappa^d}}
    {\Det[1-\tilde r_{\kappa}\tilde G_{\kappa}(\beta)]}
    \int_{\R}dx\,\frac{e^{-i\tilde s_{\kappa}x-\kappa^dx^2/2\beta\lambda}}
  {\Det[1- (e^{ix}-1)\tilde r_{\kappa}\tilde G_{\kappa}(\beta)
          (1-\tilde r_{\kappa}\tilde G_{\kappa}(\beta))^{-1}]} .
\end{equation}
Here $(\tilde s_{\kappa}, \tilde r_{\kappa})$ is the unique solution of
\begin{equation}
\begin{cases}
            \tilde r = \exp\big(\beta\mu - \beta\lambda \tilde s/\kappa^d )  \\
            \tilde s = \Tr [\tilde r\tilde G_{\kappa}(\beta)(1-\tilde r\tilde G_{\kappa}(\beta))^{-1}].
\end{cases}
\label{tilde_rs}
\end{equation}
Obviously, $\tilde r_{\kappa} \in (0, ||\tilde G_{\kappa}(\beta)||^{-1})$.
Note also that $r_{\kappa}$ and $\tilde r_{\kappa}$ satisfy the following
conditions respectively:
\begin{eqnarray}
          \frac{1}{\kappa^d}\Tr [r_{\kappa} G_{\kappa}(\beta)(1- r_{\kappa}G_{\kappa}(\beta))^{-1}]
         = \frac{\beta\mu-\log r_{\kappa}}{\beta\lambda},
\label{r}\\
        \frac{1}{\kappa^d}\Tr [\tilde r_{\kappa} \tilde G_{\kappa}(\beta)
         (1-\tilde r_{\kappa} \tilde G_{\kappa}(\beta))^{-1}]
         = \frac{\beta\mu-\log \tilde r_{\kappa}}{\beta\lambda}.
\label{rt}
\end{eqnarray}

Since by definition (\ref{G-f}) one obviously gets:
$\tilde G_{\kappa}(\beta) \leqslant G_{\kappa}(\beta)$, the operator $\tilde G_{\kappa}(\beta)$
also belongs to the  trace-class $\mathcal{C}_1 (\mathfrak{H})$.
We put the eigenvalues of $\tilde G_{\kappa}(\beta)$ in the decreasing order
\[
         \tilde g_0^{(\kappa)} = ||\tilde G_{\kappa}(\beta)|| \geqslant
          \tilde g_1^{(\kappa)} \geqslant \cdots
\]
Then, we have $ g_j^{(\kappa)} \geqslant \tilde g_j^{(\kappa)}
 \quad ( j= 0, 1,2, \cdots )$ by the \textit{min-max principle}.

\subsection{ Approximations of One-Particle Gibbs Semigroups}
Here we  establish some relations between Gibbs semigroup $\{G_{\kappa}(\beta)\}_{\beta \geq 0}$
and the heat semigroup $\{G(\beta)\}_{\beta \geq 0}$.
Let $P_{\kappa}$ be the orthogonal projection on $\mathfrak{H}$ onto its one-dimensional
subspace spanned by the vector $\Omega_{\kappa}$, and put $ Q_{\kappa}: = I - P_{\kappa}$.
\begin{lem}
For any $ r \in ( 0, 1)$,
\begin{eqnarray}
      ||\sqrt{1-e^{-f}}\,\big[rG_{\kappa}(\beta)(1-rG_{\kappa}(\beta))^{-1}
        -rG(\beta)(1-rG(\beta))^{-1}\big] \sqrt{1-e^{-f}}||_1 &\to& 0, \qquad
\label{rG}  \\
  ||\sqrt{1-e^{-f}}Q_{\kappa}G_{\kappa}(\beta)Q_{\kappa}
          (1-Q_{\kappa}G_{\kappa}(\beta)Q_{\kappa})^{-1}\sqrt{1-e^{-f}}
        -K_f||_1 &\to& 0
\label{QG}
\end{eqnarray}
hold in the limit $\kappa \to \infty$, where $|| \, \cdot \, ||_1$ stands for the trace norm
in $\mathcal{C}_1 (L^2(\R^d))$.
\end{lem}
{\sl Proof :} First, we show the estimates
\begin{eqnarray}
   |G_{\kappa}^n(\beta; x,y) -G^n(\beta; x,y)| \leqslant \frac{A'}{\kappa n^{d/2-1}}
   \Big(1+\frac{|x|^2+|y|^2}{\kappa}\Big) & \mbox{ if } &
        n\beta/\kappa \leqslant 1,
\label{G-G}\\
    |G_{\kappa}^n(\beta; x, y) - \Omega_{\kappa}(x)\Omega_{\kappa}(y)|
   \leqslant \frac{B'e^{-n\beta/2\kappa}}{\kappa^{d/2}}
    \Big(1+\frac{|x-y|^2}{\kappa}\Big) & \mbox{ if } &
       n\beta/\kappa \geqslant 1,
\label{G-O}
\end{eqnarray}
where $ A'$ and $B'$ depend only on $d$ and $\beta$.
In fact, by the \textit{Mehler's formula} one gets for $n\beta/\kappa \leqslant 1$ the estimate:
\begin{eqnarray*}
  &&|G_{\kappa}^n(\beta; x,y) - G^n(\beta; x,y)|
\\
   &\leqslant& \frac{1}{(2\pi n\beta)^{d/2}}
    \bigg|\bigg(\frac{2n\beta/\kappa}{1-e^{-2n\beta/\kappa}}\bigg)^{d/2}-1\bigg|
   \exp\Big(-\frac{\tanh (n\beta/2\kappa)}{2\kappa}(|x|^2+|y|^2)
        - \frac{|x-y|^2}{2\kappa\sinh (n\beta/\kappa)}\Big)
\\
            &+& \frac{1}{(2\pi n\beta)^{d/2}}\Big|
    \exp\Big(-\frac{\tanh (n\beta/2\kappa)}{2\kappa}(|x|^2+|y|^2)\Big) -1\Big|
     \exp\Big(- \frac{|x-y|^2}{2\kappa\sinh (n\beta/\kappa)}\Big)
\\
          &+& \frac{1}{(2\pi n\beta)^{d/2}}\bigg|
     1- \exp\bigg(-\Big( \frac{\sinh (n\beta/\kappa)}{(n\beta/\kappa)} -1\Big)
         \frac{|x-y|^2}{2\kappa\sinh (n\beta/\kappa)}\bigg)\bigg|
          \exp\Big(- \frac{|x-y|^2}{2\kappa\sinh(n\beta/\kappa)}\Big)
\\
   &\leqslant& \frac{1}{(2\pi n\beta)^{d/2}}\Big(
      \frac{An\beta}{\kappa} +  \frac{|x|^2+|y|^2}{2\kappa}\frac{n\beta}{2\kappa}
   +\Big(\frac{n\beta}{\kappa}\Big)^2\frac{|x-y|^2}{2\kappa\sinh (n\beta/\kappa)}\Big)
         \exp\Big(-\frac{|x-y|^2}{2\kappa\sinh(n\beta/\kappa)}\Big)
\\
  &\leqslant& \frac{A'}{\kappa n^{d/2-1}}\Big(1+\frac{|x|^2+|y|^2}{\kappa}\Big) \ .
\end{eqnarray*}
Here we have used (\ref{kei2}) for the first term, (\ref{exp1}, \ref{th1})
for the second term and (\ref{exp1}, \ref{sh1}) for the third term at the second
inequality and (\ref{sh1}) at the last inequality.

On the other hand for $ n\beta/\kappa \geqslant 1$, we obtain:
\begin{eqnarray*}
  &&|G_{\kappa}^n(\beta;x,y) - \Omega_{\kappa}(x)\Omega_{\kappa}(y)|
\\
   &\leqslant& \frac{1}{(\pi\kappa)^{d/2}}
    \bigg|\bigg(\frac{1}{1-e^{-2n\beta/\kappa}}\bigg)^{d/2}-1\bigg|
   \exp\Big(-\frac{\tanh (n\beta/2\kappa)}{2\kappa}(|x|^2+|y|^2)
        - \frac{|x-y|^2}{2\kappa\sinh (n\beta/\kappa)}\Big)
\\
            &+& \frac{1}{(\pi\kappa)^{d/2}}\Big|
    \exp\Big(-\frac{\tanh (n\beta/2\kappa)}{2\kappa}(|x|^2+|y|^2)\Big)
          -\exp\Big(-\frac{|x|^2+|y|^2}{2\kappa}\Big)\Big|
     \exp\Big(- \frac{|x-y|^2}{2\kappa\sinh (n\beta/\kappa)}\Big)
\\
          &+& \frac{1}{(\pi\kappa)^{d/2}}
          \exp\Big(-\frac{|x|^2+|y|^2}{2\kappa}\Big)
       \Big|\exp\Big(- \frac{|x-y|^2}{2\kappa\sinh (n\beta/\kappa)}\Big)
          -1\Big|
\\
   &\leqslant& \frac{1}{(\pi\kappa)^{d/2}}\Big(
      Be^{-2n\beta/\kappa} +  \frac{1}{e}\Big(\coth\Big(\frac{n\beta}{2\kappa}\Big)
     -1\Big)
   +\frac{|x-y|^2}{2\kappa\sinh (n\beta/\kappa)}\Big)
\\
  &\leqslant& \frac{B'e^{-n\beta/2\kappa}}{\kappa^{d/2}}
         \Big(1+\frac{|x-y|^2}{\kappa}\Big) \ ,
\end{eqnarray*}
where we have used (\ref{kei3}) for the first term, (\ref{exp2},\ref{th2})
for the second term and  (\ref{exp1},\ref{sh2}) for the third term.

Now, let us show the second limit (\ref{QG}), notice that the inequality:
\begin{eqnarray*}
          &&  \sum_{n=1}^{\infty}|G_{\kappa}^n(\beta; x,y) -G^n(\beta; x,y)-
          \Omega^{\kappa}_0(x)\Omega_{0, \kappa}(y)|
\\
   & \leqslant & \sum_{n=1}^{\lceil \kappa/\beta \rceil}
     \bigg(\frac{A'}{\kappa n^{d/2-1}} \Big(1+\frac{|x|^2+|y|^2}{\kappa}\Big)
     + \frac{1}{(\pi\kappa)^{d/2}}\bigg)
\\
      & + & \sum_{n=\lceil \kappa/\beta \rceil +1}^{\infty}
     \bigg(\frac{B'e^{-n\beta/2\kappa}}{\kappa^{d/2}}
    \Big(1+\frac{|x-y|^2}{\kappa}\Big)
     + \frac{1}{(2\pi n\beta)^{d/2}}\bigg)
\\
   & \leqslant & C'(\kappa^{1-d/2}\vee \kappa^{-1}\log\kappa ) \Big(1+\frac{|x|^2+|y|^2}{\kappa}\Big),
\end{eqnarray*}
holds for $C'$, which depends only on $d$ and $\beta$.
The \textit{integer part} is denoted by $\lceil \cdot \rceil$.
Here we used the estimates (\ref{G-G}), (\ref{G-O}) and
\[
  \sum_{n=1}^N \frac{1}{n^{d/2-1}} =
\begin{cases}
            O(1)         & (d/2 >2)\\
            O(N^{2-d/2}) & (1<) d/2 <2 \\
            O(\log N)     &  d/2 =2
\end{cases}
\]
and so on.
Now put
\[
         A^{(\kappa)} :=  \sqrt{1-e^{-f}} Q_{\kappa}G_{\kappa}(\beta)Q_{\kappa}
         (1- Q_{\kappa}G_{\kappa}(\beta)Q_{\kappa})^{-1} \sqrt{1-e^{-f}}.
\]
Then, since $||Q_{\kappa}G_{\kappa}(\beta)Q_{\kappa}|| = e^{-\beta/\kappa} <1$, one gets in the limit
$N\rightarrow\infty$  the operator-norm convergence:
\begin{equation}
   A_N^{(\kappa)}  =  \sqrt{1-e^{-f}} \sum_{n=1}^N
       Q_{\kappa}G_{\kappa}(\beta)^nQ_{\kappa} \sqrt{1-e^{-f}} \to A^{(\kappa)} \geq 0 \ .
\label{ANL}
\end{equation}
Recall that Theorem 3.1(i) and Proposition 2.3(i) of \cite{2} yield the
\textit{strong} convergence:
\begin{equation}
K_N  =   \sqrt{1-e^{-f}} \sum_{n=1}^N G(\beta)^n \sqrt{1-e^{-f}} \to K_f \geq 0
\label{KNL}
\end{equation}
for $N\rightarrow\infty$.
Moreover, we also have the following estimate for the operator norm.
\[
     ||A^{(\kappa)}_N - K_N|| = \sup_{||\phi||=1}|(\phi,
        \sqrt{1-e^{-f}} \sum_{n=1}^N
       (Q_{\kappa}G_{\kappa}(\beta)^nQ_{\kappa} - G(\beta)^n) \sqrt{1-e^{-f}} \, \phi)|
\]
\[
        \leqslant \sup_{||\phi||=1}\int_{\supp f}dx\int_{\supp f} dy
        \sqrt{1-e^{-f(x)}} \sqrt{1-e^{-f(y)}} |\phi(x)\phi(y)|
\]
\[
       \times \sum_{n=1}^{N}|G_{\kappa}^n(\beta; x,y) -G^n(\beta; x,y)-
          \Omega_{\kappa}(x)\Omega_{\kappa}(y)|
\]
\[
        \leqslant ||\sqrt{1-e^{-f}}||^2 \sup_{x, y\in\supp f}
          \sum_{n=1}^{\infty}|G_{\kappa}^n(\beta; x,y) -G^n(\beta; x,y)-
          \Omega^{\kappa}_0(x)\Omega_{0, \kappa}(y)|  \to 0
\]
for $\kappa\rightarrow\infty$ \textit{uniformly} in $N$.
Here $||\sqrt{1-e^{-f}}||$ stands for the $L^2$ norm of the function.
We have used Cauchy-Schwarz inequality at the second inequality.
The standard $3 \epsilon$-argument yields that
$ A^{(\kappa)} \to K_f $ strongly when $\kappa\rightarrow\infty$.

On the other hand, since the operators  $A^{(\kappa)}, K_f $ are \textit{non-negative},
we have for $\kappa\rightarrow\infty$ the limit:
\begin{eqnarray*}
      ||A^{(\kappa)}||_1 - ||K_f||_1 &=& \Tr A^{(\kappa)} - \Tr K_f =
      \sum_{n=1}^{\infty} (\phi_n, (A^{(\kappa)} -  K_f)\phi_n)
\\
  &=& \sum_{n=1}^{\infty}\sum_{l=1}^{\infty} (\phi_n, \sqrt{1-e^{-f}}(Q_{\kappa}G_{\kappa}(\beta)^l
    Q_{\kappa}
    - G(\beta)^l)\sqrt{1-e^{-f}}\phi_n)
\\
     &=& \sum_{l=1}^{\infty}\sum_{n=1}^{\infty} (\phi_n, \sqrt{1-e^{-f}}
       (Q_{\kappa}G_{\kappa}(\beta)^lQ_{\kappa} - G(\beta)^l)\sqrt{1-e^{-f}}\phi_n)
\\
    &=& \sum_{l=1}^{\infty}\Tr \sqrt{1-e^{-f}}
    (Q_{\kappa}G_{\kappa}(\beta)^lQ_{\kappa} - G(\beta)^l)\sqrt{1-e^{-f}}
\\
   &=& \sum_{l=1}^{\infty}\int_{\supp f}(1-e^{-f(x)})(G_{\kappa}^l(\beta; x, x)
    - G^l(\beta; x,x) -\Omega_{\kappa}(x)^2)\, dx \to 0,
\end{eqnarray*}
where $\{\phi_n\}_{n =1} ^{\infty}$ is
{an arbitrary complete ortho-normal system} in $\mathfrak{H}$.
Note that we can exchange the order of summations over $n$ and $l$, since
\begin{equation*}
\Tr A^{(\kappa)} = \sum_{n,l} (\phi_n, \sqrt{1-e^{-f}}Q_{\kappa}G_{\kappa}(\beta)^l
Q_{\kappa}\sqrt{1-e^{-f}}\phi_n)
\end{equation*}
and
\[
    \Tr K_f =  \sum_{n,l} (\phi_n, \sqrt{1-e^{-f}} G^l(\beta)\sqrt{1-e^{-f}}\phi_n)
\]
are convergent non-negative sequences. Here we used (\ref{ANL}) and (\ref{KNL}) in the third equality above.
Thus, we get $ \lim_{\kappa \rightarrow \infty } A^{(\kappa)}= K_f $ in $\mathfrak{C}_1(\mathfrak{H})$ by
the \textit{Gr\"umm convergence} theorem, see e.g. \cite{Zag}.

Let us consider the first limit (\ref{rG}).
Since the identity $rG_{\kappa}(\beta)(1-rG_{\kappa}(\beta))^{-1} =\sum_{n=1}^{\infty}r^nG_{\kappa}(\beta)^n$
holds in the operator norm topology and $G_{\kappa}^n(\beta;x,y) \geqslant 0$
,(by virtue of  Lemma 2.2(ii) in \cite{2}) we get the representation:
\[
      \big[rG_{\kappa}(\beta)(1-rG_{\kappa}(\beta))^{-1}\big](x,y)
        =\sum_{n=1}^{\infty}r^nG_{\kappa}^n(\beta;x,y).
\]
Similarly, one gets the representation:
\[
      \big[rG(\beta)(1-rG(\beta))^{-1}\big](x,y) =\sum_{n=1}^{\infty}r^n G^n(\beta; x,y) \ .
\]
In fact the series in the right-hand side of the above representations are \textit{point-wise}
convergent because the uniform estimates
\[
    G_{\kappa}^n (\beta; x,y)\leqslant
       \Big(\frac{1}{\pi\beta}\vee\frac{2}{\pi\kappa}\Big)^{d/2},
 \quad G^n(\beta; x,y) \leqslant \Big(\frac{1}{2\pi\beta}\Big)^{d/2}
\]
hold for all $\kappa>0$, $ x, y \in \R^d $ and $ n\in\N$. Here we used (\ref{kei1}) for the first inequality.
Hence we obtain the estimate
\[
     \big|\big[rG_{\kappa}(\beta)(1-rG_{\kappa}(\beta))^{-1}\big](x,y) -
         \big[rG(\beta)(1-rG(\beta))^{-1}\big](x,y) \big| \leqslant
      \sum_{n=1}^{\infty}r^n|G_{\kappa}^n(\beta; x,y)-G^n(\beta; x,y)|
\]
\[
      \leqslant \sum_{n=1}^{\lceil\kappa/\beta\rceil}r^n\frac{A'}{\kappa n^{d/2-1}}
   \Big(1+\frac{|x|^2+|y|^2}{\kappa}\Big)
            + \sum_{n=\lceil\kappa/\beta\rceil+1}^{\infty}r^n
           \Big[\Big(\frac{1}{2\pi\beta}\Big)^{d/2}+
               \Big(\frac{1}{\pi\beta}\vee\frac{2}{\pi\kappa}\Big)^{d/2}\Big]
\]
\[
       \leqslant  \frac{A'}{\kappa (1-r)}
   \Big(1+\frac{|x|^2+|y|^2}{\kappa}\Big)
        + C\frac{r^{\lceil\kappa/\beta\rceil+1}}{1-r}
\]
which tends to zero when $\kappa \to \infty$, uniformly in $x,y \in \mathfrak{C}$ for any
compact set $\mathfrak{C}$, where
$C$ denotes a constant which depends only on $d$ and on $\beta$.

Thus,
\[
  \sup_{x,y\in \supp f}
      \big|\big[rG_{\kappa}(\beta)(1-rG_{\kappa}(\beta))^{-1}\big](x,y) -
         \big[rG(\beta)(1-rG(\beta))^{-1}\big](x,y) \big| \to 0
\]
holds.
We get the first of the announced limits (\ref{rG}) by the similar (even simpler) argument
to the second one.
This finishes the  proof of the lemma.
\hfill$\Box$

\subsection{Estimates for the Scaled Mean-Field Interaction}

In the followings, we use the notation $B_\kappa := \hat O(\kappa^{\alpha})$, which means
that there exist two numbers $c_1 \geqslant c_2 > 0$ such that
\begin{equation*}
c_1 \kappa^{\alpha} \geqslant B_\kappa \geqslant c_2 \kappa^{\alpha}.
\end{equation*}
Next, we put $W_{\kappa}: = (G_{\kappa}(\beta))^{1/2}\sqrt{1-e^{-f}} $ and define
$ D_{\kappa}: = G_{\kappa}(\beta) - \tilde G_{\kappa}(\beta) =W_{\kappa}W^*_{\kappa}$.

\begin{lem}
For large $\kappa > 0$, the following asymptotics hold:
\[
    (\Omega_{\kappa}, D_{\kappa}\Omega_{\kappa}) = ||W_{\kappa}^* \Omega_{\kappa}||^2
      = ||\sqrt{ 1 - e^{-f}} ||^2 (1+o(1))/(\pi\kappa)^{d/2},
\]
\[
      \Tr D_{\kappa} = \frac{ ||\sqrt{ 1- e^{-f}} ||^2(1+o(1))}
           {\big(\pi\kappa(1-e^{-2\beta/\kappa})\big)^{d/2}}
\]
and
\begin{eqnarray*}
g_0^{(\kappa)} - \tilde g_0^{(\kappa)} &=&
(\Omega_{\kappa}, (D_{\kappa} - D_{\kappa}Q_{\kappa}
[\tilde g_0^{(\kappa)}-Q_{\kappa}\tilde G_{\kappa}(\beta)Q_{\kappa}]^{-1}Q_{\kappa}D_{\kappa})
\Omega_{\kappa})\\
& = & \frac{1+o(1)}{(\pi\kappa)^{d/2}}(\sqrt{1-e^{-f}},
[1 + W_{\kappa}^*Q_{\kappa}[1-Q_{\kappa}G_{\kappa}(\beta)Q_{\kappa}]^{-1}
Q_{\kappa}W_{\kappa}]^{-1}
\sqrt{1-e^{-f}})\\
& = & \hat O(\kappa^{-d/2}) \ .
\end{eqnarray*}
\label{gt}
\end{lem}
{\sl Proof:} For simplicity of notation we suppress everywhere below the index $\kappa$ in
$ g_j^{(\kappa)}, \tilde g_j^{(\kappa)}, \Omega_{(\kappa)} $ and in  $Q_{\kappa}$.

Note that the \textit{first} equality is a straightforward consequence of
definitions (\ref{Harmon-Ground}),
(\ref{G-f}). The \textit{second} equality can be derived directly from the
\textit{Mehler's formula}.

Now by the \textit{min-max} principle, for $d>2$ and $\kappa$ large enough,
we obtain from the value  $g_1 = \exp(-\beta/\kappa)$ the following estimates:
\begin{eqnarray}
g_0&=& 1 \geqslant \tilde g_0 \geqslant (\Omega, \tilde G_{\kappa}(\beta) \Omega)
= 1 - (\Omega, D_{\kappa} \Omega) \\ \nonumber
&=&1 - \hat O(\kappa^{-d/2})
> g_1 = 1 - \hat O(\kappa^{-1}) \geqslant \tilde g_1 \ . \label{V}
\end{eqnarray}
Hence the eigenspace of the operator $\tilde G_{\kappa}(\beta)$ for its largest eigenvalue  $ \tilde g_0$
is \textit{one-dimensional}.
Let $\tilde \Omega $ be the normalized eigenfunction corresponding to $\tilde g_0$ and let
$\tilde\Omega = a\Omega + \Omega'$ ,
with $(\Omega, \Omega') = 0$.
Then $ \tilde G_{\kappa}(\beta)\tilde\Omega = \tilde g_0\tilde\Omega $
yields
\begin{equation*}
a\tilde G_{\kappa}(\beta)\Omega + \tilde G_{\kappa}(\beta)\Omega' = a\tilde g_0\Omega
+ \tilde g_0\Omega' \ .
\end{equation*}
Applying to this relation orthogonal projector $P$ (on $\Omega$) and
$Q= I - P$, we obtain:
\begin{eqnarray*}
      ag_0 - a (\Omega, D_{\kappa}\Omega) - (\Omega, D_{\kappa}\Omega')
      &=& a\,\tilde g_0 \ ,     \\
      -aQD_{\kappa}\Omega + Q \tilde G_{\kappa}(\beta)Q\Omega' &=& \tilde g_0\,\Omega' \ .
\end{eqnarray*}
Since $ Q\tilde G_{\kappa}(\beta)Q \leqslant Q G_{\kappa}(\beta)Q \leqslant g_1 < \tilde g_0$,
the operator $\tilde g_0 - Q\tilde G_{\kappa}(\beta)Q $ is positive and invertible.
It follows from the second identity that
\begin{equation}
       \Omega'  =  -a[\tilde g_0 -Q\tilde G_{\kappa}(\beta)Q]^{-1}QD_{\kappa}\Omega.
 \label{Omega'}
\end{equation}
Together with the above first identity,
\begin{eqnarray}
      g_0 - \tilde g_0
      & = & (\Omega, (D_{\kappa} - D_{\kappa}Q[\tilde g_0-Q\tilde G_{\kappa}(\beta)Q]^{-1}
             QD_{\kappa})\Omega)
  \notag\\
           & = & (W_{\kappa}^*\Omega,(1 -  W_{\kappa}^*Q
           [\tilde g_0-Q\tilde G_{\kappa}(\beta)Q]^{-1}QW_{\kappa})W_{\kappa}^*\Omega)
\label{i}
\end{eqnarray}
follows.

For brevity, we put
\[
X' := W_{\kappa}^*Q[\tilde g_0-QG_{\kappa}(\beta)Q]^{-1}QW_{\kappa} \ , \quad
X  := W_{\kappa}^*Q[1 - QG_{\kappa}(\beta)Q]^{-1}QW_{\kappa}
\]
and
\[
  \tilde X := W_{\kappa}^*Q[\tilde g_0-Q\tilde G_{\kappa}(\beta)Q]^{-1}QW_{\kappa} \ .
\]
Then we get
\[
  \tilde X - X' = - \tilde X X',
\]
and hence
\begin{equation}
    \tilde X =  X'(1+X')^{-1} \quad \mbox{and} \quad 1-\tilde X = (1+X')^{-1}.
\label{XX'}
\end{equation}
By definition of $W_{\kappa}$ and (\ref{Harmon-Ground}) one gets for large $\kappa$ the asymptotic:
\begin{equation*}
W_{\kappa}^*\Omega = \sqrt{1-e^{-f}}(\pi\kappa)^{-d/4}(1+O(\kappa^{-1})) \ .
\end{equation*}
By virtue of (\ref{i}) it implies the representation:
\begin{equation}
      g_0 - \tilde g_0
             =  (\pi\kappa)^{-d/2}(\sqrt{1-e^{-f}},(1 + X')^{-1} \sqrt{1-e^{-f}})
            (1+O(\kappa^{-1})) \ .
\label{f}
\end{equation}

Now, we want to replace in the right hand side of this representation the operator $X'$ by $X$ .
Note that (\ref{QG}) yields $||X - K_f||\leqslant ||X - K_f||_1 = o(1)$.
Then by $1 - \tilde g_0 =  O({\kappa}^{-d/2}), \; \tilde g_0 - g_1 = \hat O({\kappa}^{-1}), \
1-g_1 = \hat O(\kappa^{-1}) $
and $ ||W_{\kappa}^*\Omega|| = \hat O(\kappa^{-d/4})$,
we find that
\begin{eqnarray*}
    || X' - X || & = & (1-\tilde g_0)||W_{\kappa}^*Q[\tilde g_0-QG_{\kappa}(\beta)Q]^{-1}
       [1-QG_{\kappa}(\beta)Q]^{-1}QW_{\kappa}||
\\
       & \leqslant & \frac{1-\tilde g_0}{\tilde g_0 - g_1}
    || W_{\kappa}^*Q[1-QG_{\kappa}(\beta)Q]^{-1}QW_{\kappa}||
\\
     & = & O(\kappa^{1-d/2})||X|| =  O(\kappa^{1-d/2})||K_f|| \ ,
\end{eqnarray*}
which implies $ || X' || \leqslant ||K_f||(1+o(1))$.
Hence (\ref{f}) yields
\[
   (\pi\kappa)^{d/2}(g_0 - \tilde g_0) = (\sqrt{1-e^{-f}}, ( 1+ X' )^{-1} \sqrt{1-e^{-f}})
(1 + O(\kappa^{-1}))
   \geqslant \frac{||\sqrt{1-e^{-f}}||^2}{1+||K_f||}(1+o(1)).
\]

Since the  upper bound $(\pi\kappa)^{d/2}(g_0 - \tilde g_0) \leqslant (\pi\kappa)^{d/2}(\Omega, D_{\kappa}
\Omega) \leqslant ||\sqrt{1-e^{-f}}||^2$ is obvious, we get desired estimate
$ g_0 - \tilde g_0 = \hat O(\kappa^{-d/2})$.
 Notice that the proof of the equality
\begin{equation*}
g_0 - \tilde g_0= \frac{1+o(1)}{(\pi\kappa)^{d/2}}(\sqrt{1-e^{-f}}, [1 + X]^{-1} \sqrt{1-e^{-f}})
\end{equation*}
follows from (\ref{f}) and the estimate:
\[
     |(\sqrt{1-e^{-f}}, ( 1+ X' )^{-1} \sqrt{1-e^{-f}})
     - (\sqrt{1-e^{-f}}, ( 1+ X )^{-1} \sqrt{1-e^{-f}})|
\]
\[
     \leqslant ||\sqrt{1-e^{-f}}||^2 ||(1 + X)^{-1}||\,||(1 + X')^{-1}||\,||X-X'||
     = o(1) \ .
\]
Here $\|\sqrt{1-e^{-f}}\|$ stands for the norm in $ \mathfrak{H} =L^2(\R^d)$,
while the other $\|\cdot\|$ for the operator norm on $ \mathfrak{H}$.
This remark finishes the proof of the lemma.
\hfill$\Box$

\subsection{Evidence of Two Thermodynamic Regimes}\label{2-regimes}
Now we return to the conditions (\ref{r}).
We need the behavior of $r_{\kappa}$ and $\tilde r_{\kappa}$ to prove the main theorem.
Here we consider the behavior of $r_{\kappa}$, which classifies the phase separation.
That of $\tilde r_{\kappa}$ is postponed to Section 4.
\begin{prop}\label{cases}
{\rm{(a)}}  $ \{ r_{\kappa} \} $ converges to  $ r_* \in (0, 1)$ for $\kappa \to \infty$,
where $r_*$ is the unique solution of
\begin{equation}
  \frac{\mu}{\lambda} = \frac{\log r_*}{\beta\lambda}
     + \frac{1}{\beta^d}\int_{[0,\infty)^d}\frac{r_*dp}{e^{|p|_1} -r_*} \ ,
\label{r*}
\end{equation}
if and only if $ \beta^d\mu < \zeta(d)\lambda $ .\\
{\rm{(b)}}  $ {\kappa}^d(1-r_{\kappa}) \longrightarrow \beta^d\lambda/(\beta^d\mu - \zeta(d)\lambda)$,
and hence  $\lim_{\kappa\to\infty} r_{\kappa} =1$, if and only if $ \beta^d\mu > \zeta(d)\lambda  $.\\
{\rm{(c)}} $\lim_{\kappa \to \infty} r_{\kappa}=1$
and
      $ \kappa^d(1-r_{\kappa}) \longrightarrow +\infty$ ,
if and only if $ \beta^d\mu = \zeta(d)\lambda $.
\end{prop}

To this end let us introduce the notation:
\[
            \Box_n^{(\kappa)} : \ = \ \frac{\beta}{\kappa}(n + [0, 1)^d)
         \qquad \mbox{for} \quad n \in \Z_+^d,
\]
and define for $r \in [0, 1], \; \nu=1, 2 $ and $\kappa \in [1,\infty)$,
the functions $a_{\nu}( \,\cdot \,;r), a_{\nu}^{(\kappa)}(\,\cdot\,;r)$ on $[0,\infty)^d$ by
\[
    a_{\nu}(p; r) := \frac{re^{-|p|_1}}{(1 - re^{-|p|_1})^{\nu}}
\]
and by
\[
     a_{\nu}^{(\kappa)}(p; r) :=
     \begin{cases}
     0 & \mbox{ if } \quad p\in \Box_0^{(\kappa)}
      \\
     a_{\nu}(\beta n/\kappa; r) & \mbox{ if } \quad
     p\in \Box_n^{(\kappa)}  \quad \mbox{for} \quad n \in \Z_+^d -\{0\}.
     \end{cases}
\]
It is easy to show the following fact. (See \cite{2} for detail.)
\begin{lem}\label{a}
There exists a constant $c$ which depends only on $d>2$ and $\beta>0$ such that
\[
    0 \leqslant a_{\nu}^{(\kappa)}(p;r)\leqslant a_{\nu}(c \, p; 1) \in L^1([0,\infty)^d)
\]
holds for $ r \in [0,1] $ and $ \nu = 1, 2 $.
\end{lem}
\begin{remark} \label{REMa}
If a series $\{r_{\kappa}\} \subset [0,1]$ converges to $ r_0 \in [0,1]$,
$a_{\nu}^{(\kappa)}( \cdot, r_{\kappa}) \to a_{\nu}(\cdot, r_0) $ holds $ a.e.$.
Then the lemma and the dominated convergence theorem yield
\[
     \frac{\Tr [r_{\kappa}Q_{\kappa}G_{\kappa}(\beta)Q_{\kappa}
(1-r_{\kappa}Q_{\kappa}G_{\kappa}(\beta)Q_{\kappa})^{-\nu}]}
      {\kappa^d}
    = \frac{1}{\kappa^d}\sum_{j=1}^{\infty}
        \frac{r_{\kappa}g_j^{(\kappa)}}{(1-r_{\kappa}g_j^{(\kappa)})^{\nu}}
\]
\[
    = \Big(\frac{1}{\beta^d}\Big) \int_{[ 0, \infty)^d}a_{\nu}^{(\kappa)}( p; r_{\kappa})\, dp
    \to \Big(\frac{1}{\beta^d}\Big) \int_{[ 0, \infty)^d}a_{\nu}( p; r_0)\, dp
\]
as $\kappa \to \infty$.
\end{remark}
{\sl Proof of Proposition \ref{cases}:} If $r_{\kappa} \to r_*\in [ 0, 1)$, then by (\ref{r})
and the above remark one gets:
\[
      \frac{\mu}{\lambda} =
      \frac{\log r_{\kappa}}{\beta\lambda} + \frac{1}{\beta^d}
      \int_{[0,\infty)^d}a_{1}^{(\kappa)}(p;r_{\kappa})\, dp
          +\frac{r_{\kappa}}{(1-r_{\kappa})\kappa^d}
\]
\[
      \longrightarrow \frac{\log r_*}{\beta\lambda} + \frac{1}{\beta^d}
       \int_{[0,\infty)^d}a_{1}(p;r_*)\,dp <
          \frac{1}{\beta^d} \int_{[0,\infty)^d}a_{1}(p;1)\,dp
              =\frac{\zeta(d)}{\beta^d} \ .
\]
Since equality holds at the limit actually, (\ref{r*}) follows.
Similarly, if $r_{\kappa} \to 1$ and $ \kappa^d(1-r_{\kappa}) \to \infty$, then one obtains:
\[
      \frac{\mu}{\lambda} =
      \frac{\log r_{\kappa}}{\beta\lambda} + \frac{1}{\beta^d}
      \int_{[0,\infty)^d}a_{1}^{(\kappa)}(p;r_{\kappa})\, dp
          +\frac{r_{\kappa}}{(1-r_{\kappa})\kappa^d}
\]
\[
        \longrightarrow  \frac{1}{\beta^d}
       \int_{[0,\infty)^d}a_{1}(p;1) = \frac{\zeta(d)}{\beta^d}.
\]
Finally, if $ \kappa^d(1-r_{\kappa}) \to \alpha >0$,
we get
\[
      \frac{\mu}{\lambda} =\frac{1}{\alpha} + \frac{\zeta(d)}{\beta^d} > \frac{\zeta(d)}{\beta^d}
   \quad \mbox{ and } \quad \alpha =\frac{\beta^d\lambda}{\beta^d\mu - \zeta(d)\lambda}.
\]
On the other hand, if $\{r_{\kappa}\}$ does not converge, by compactness we can take
two convergent subsequences  $\{r_{\kappa_i}\} \quad (i=1,2)$ with different limits
$ 0 <r^{(1)} < r^{(2)} <1$. Then above arguments yield
\[
      \frac{\log r^{(1)}}{\beta\lambda} + \frac{1}{\beta^d}
       \int_{[0,\infty)^d}a_{1}(p;r^{(1)})\,dp
    = \frac{\log r^{(2)}}{\beta\lambda} + \frac{1}{\beta^d}
       \int_{[0,\infty)^d}a_{1}(p;r^{(2)})\,dp \ ,
\]
which contradicts to the strict monotonicity of the function
\begin{equation}
          h(r) = \frac{\log r}{\beta\lambda} + \frac{1}{\beta^d}
       \int_{[0,\infty)^d}a_{1}(p;r)\,dp \ .
\label{hr}
\end{equation}
Similar arguments are valid for the cases $r^{(1)} = 0 $ or $r^{(2)} = 1 $.
If  $r_{\kappa}\to 1$ but  $ \kappa^d(1-r_{\kappa}) $ does not converge,
we again get a contradiction. \hfill$\Box$
\section{ Proof of Theorem \ref{thmA}}
\subsection{ The case {$ \mu \leq \mu_c(\beta)$} (normal phase). }
Let us recall that for the weak convergence of random point fields, it is enough to
prove the convergence of the generating functionals. Therefore, we have to evaluate the integral in (\ref{pf2}).
We begin with estimates of the Fredholm determinant in the integrand.
For all values of $x$, we have
\begin{eqnarray}
\lefteqn{
 \left|{\rm Det}\left(1-(e^{ix}-1)\frac{r_{\kappa}G_{\kappa}(\beta)}
   {1-r_{\kappa}G_{\kappa}(\beta)}\right)\right|^{2}
        } &&
\notag \\
&=& {\rm Det}\left[\left(1-(e^{ix}-1)\frac{r_{\kappa}G_{\kappa}(\beta)}
   {1-r_{\kappa}G_{\kappa}(\beta)}\right)
         \left(1-(e^{-ix}-1)\frac{r_{\kappa}G_{\kappa}(\beta)}
   {1-r_{\kappa}G_{\kappa}(\beta)}\right)\right]
\notag \\
&=&  \Det \left[1+4\sin^2\frac{x}{2}\frac{r_{\kappa}G_{\kappa}(\beta)}
      {(1-r_{\kappa}G_{\kappa}(\beta))^2}\right] \geqslant 1.
\label{Det1}
\end{eqnarray}
Set $\alpha \in ( d/3, d/2)$.
Then for $ |x| \leqslant \kappa^{-\alpha}$, we have
\begin{equation}
  \Det \Big[ 1-(e^{ix}-1)\frac{r_{\kappa}G_{\kappa}(\beta)}{1-r_{\kappa}G_{\kappa}(\beta)}\Big]^{-1}
          = \exp\Big[ix\Tr\frac{r_{\kappa}G_{\kappa}(\beta)}{1-r_{\kappa}G_{\kappa}(\beta)}
    -\frac{x^2}{2}\Tr\frac{r_{\kappa}G_{\kappa}(\beta)}{(1-r_{\kappa}G_{\kappa}(\beta))^2}\Big]
    \big(1+O(\kappa^{d-3\alpha})\big)
\label{det2}
\end{equation}
In fact,
\[
  \mbox{ log of the l.h.s. } = -\sum_{n=0}^{\infty}\log
    \Big(1-(e^{ix}-1)\frac{r_{\kappa}g_n^{(\kappa)}}{1-r_{\kappa}g_n^{(\kappa)}}\Big)
\]
\[
    = (e^{ix}-1)\Tr \frac{r_{\kappa}G_{\kappa}(\beta)}{1-r_{\kappa}G_{\kappa}(\beta)}
     + \frac{(e^{ix}-1)^2}{2}\Tr \Big(\frac{r_{\kappa}G_{\kappa}(\beta)}
           {1-r_{\kappa}G_{\kappa}(\beta)}\Big)^2 + R_1,
\]
where $R_1= O(\kappa^{d-3\alpha})$ since $ r_{\kappa} \to r_* \in ( 0, 1)$ and
\[
    \Tr \Big[\frac{r_{\kappa}G_{\kappa}(\beta)}{1-r_{\kappa}G_{\kappa}(\beta)}\Big]^{\ell}
     \leqslant \Big(\frac{r_{\kappa}g_0^{(\kappa)}}{1-r_{\kappa}g_0^{(\kappa)}}\Big)^{\ell-1}
      \Tr \frac{r_{\kappa}G_{\kappa}(\beta)}{1-r_{\kappa}G_{\kappa}(\beta)}
         = O(\kappa^d).
\]
For the last equality, we recall (\ref{r}) and Proposition \ref{cases}(a).

Similarly, we have
\[
     (e^{ix}-1)\Tr \frac{r_{\kappa}G_{\kappa}(\beta)}{1-r_{\kappa}G_{\kappa}(\beta)}
       =\Big(ix-\frac{x^2}{2}\Big)\Tr \frac{r_{\kappa}G_{\kappa}(\beta)}
        {1-r_{\kappa}G_{\kappa}(\beta)} + O(\kappa^{d-3\alpha})
\]
and
\[
    \frac{(e^{ix}-1)^2}{2}\Tr \Big(\frac{r_{\kappa}G_{\kappa}(\beta)}
           {1-r_{\kappa}G_{\kappa}(\beta)}\Big)^2
    = -\frac{x^2}{2}\Tr \Big(\frac{r_{\kappa}G_{\kappa}(\beta)}
           {1-r_{\kappa}G_{\kappa}(\beta)}\Big)^2 + O(\kappa^{d-3\alpha}).
\]
Thus we get (\ref{det2}) and the following lemma.
\begin{lem}
\[
     \Xi_{\kappa}( \beta,\mu) = \frac{e^{\kappa^d(\beta\mu-\log r_{\kappa})^2/2\beta\lambda}
         \big( 1+ O(\kappa^{d-3\alpha})\big)}
    {\sqrt{1+\beta\lambda\kappa^{-d}
        \Tr [r_{\kappa}G_{\kappa}(\beta)(1-r_{\kappa}G_{\kappa}(\beta))^{-2}]}
        \Det(1-r_{\kappa}G_{\kappa}(\beta))}
\]
\label{Xi1}
\end{lem}
{\sl Proof : } From the above estimates and (\ref{rs}), we have
\[
\int_{\R}dx\,\frac{e^{-is_{\kappa}x-\kappa^dx^2/2\beta\lambda}}
  {\Det[1- (e^{ix}-1)r_{\kappa}G_{\kappa}(\beta)(1-r_{\kappa}G_{\kappa}(\beta))^{-1}]}
\]
\[
    = \int_{-\kappa^{-\alpha}}^{\kappa^{-\alpha}}dx\, e^{-\kappa^dx^2/2\beta\lambda
           -is_{\kappa}x + ix\Tr [r_{\kappa}G_{\kappa}(\beta)
       (1-r_{\kappa}G_{\kappa}(\beta))^{-1}]
       -x^2\Tr [r_{\kappa}G_{\kappa}(\beta)
       (1-r_{\kappa}G_{\kappa}(\beta))^{-2}]/2}\big(1+O(\kappa^{d-3\alpha})\big) + R_2
\]
\[
    = \int_{-\infty}^{\infty}dx\, \exp\Big[-\Big(\frac{\kappa^d}{\beta\lambda}
      +\Tr \frac{r_{\kappa}G_{\kappa}(\beta)}{(1-r_{\kappa}G_{\kappa}(\beta))^2}\Big)
    \frac{x^2}{2}\Big]\,(1+O(\kappa^{d-3\alpha})\big) +R_2 + R_3,
\]
where
\[
     R_2 = \int_{|x|>\kappa^{-\alpha}}\frac{e^{-is_{\kappa}x-\kappa^dx^2/2\beta\lambda}}
  {\Det[1- (e^{ix}-1)r_{\kappa}G_{\kappa}(\beta)(1-r_{\kappa}G_{\kappa}(\beta))^{-1}]}
    = O(e^{-\tilde c\kappa^{d-2\alpha}})
\]
for some $ \tilde c > 0 $, thanks to (\ref{Det1}),
and
\[
   R_3 = - \int_{|x|>\kappa^{-\alpha}}dx\, \exp\Big[-\Big(\frac{\kappa^d}{\beta\lambda}
      +\Tr \frac{r_{\kappa}G_{\kappa}(\beta)}{(1-r_{\kappa}G_{\kappa}(\beta))^{2}}\Big)
    \frac{x^2}{2}\Big]\,(1+O(\kappa^{d-3\alpha})\big)
\]
\[
     =  O(e^{-\tilde c'\kappa^{d-2\alpha}})
\]
for some $ \tilde c' > 0$.
Then, the lemma follows from (\ref{pf2}) and (\ref{rs}).
\hfill$\Box$

For $\tilde \Xi_{\kappa}(\beta, \mu)$, we have the following asymptotics:
\begin{lem}
\[
     \tilde \Xi_{\kappa} (\beta, \mu) = \frac{e^{\kappa^d(\beta\mu-\log \tilde r_{\kappa})^2/2\beta\lambda}
         \big( 1+ O(\kappa^{d-3\alpha})\big)}
    {\sqrt{1+\beta\lambda\kappa^{-d}
        \Tr [\tilde r_{\kappa}\tilde G_{\kappa}(\beta)(1 - \tilde r_{\kappa} \tilde G_{\kappa}(\beta))^{-2}]}
        \Det(1- \tilde r_{\kappa}\tilde G_{\kappa}(\beta))}
\]
\label{tilde_Xi1}
\end{lem}
To show the formula, we note that "tilded" quantities are close to corresponding
"untilded" ones.
In fact, the following asymptotics is established.
Then it is obvious to get Lemma \ref{tilde_Xi1} by a similar argument.
\begin{lem} For large $\kappa$ one gets:
\begin{eqnarray*}
    (i)& \hskip1cm &     \tilde r_{\kappa} \geqslant r_{\kappa},
\\
  (ii)& \qquad& \tilde r_{\kappa} -r_{\kappa} = O(\kappa^{-d}),
\\
  (iii)& \qquad & \Tr\Big[\frac{r_{\kappa}G_{\kappa}(\beta)}{1- r_{\kappa}G_{\kappa}(\beta)}\Big]
           = \hat O(\kappa^d)  ,
    \qquad   \Tr\Big[\frac{\tilde r_{\kappa}\tilde G_{\kappa}(\beta)}
         { 1- \tilde r_{\kappa}\tilde G_{\kappa}(\beta)}\Big] = \hat O(\kappa^d) ,
\\
  (iv)& \qquad & \Tr\Big[\frac{r_{\kappa}G_{\kappa}(\beta)}{(1- r_{\kappa}G_{\kappa}(\beta))^2}\Big] -
   \Tr\Big[\frac{\tilde r_{\kappa}\tilde G_{\kappa}(\beta)}
         { (1- \tilde r_{\kappa}\tilde G_{\kappa}(\beta))^2}\Big] =  O(1) .
\end{eqnarray*}
\label{rtilde_r}
\end{lem}
{\sl Proof : } (i)  Let $ h_{\kappa}$ be the functions on $(0, 1)$ defined by
\[
    h_{\kappa}(r) = \frac{\log r}{\beta\lambda} + \frac{1}{\kappa^d}
        \Tr\frac{rG_{\kappa}(\beta)}{1-rG_{\kappa}(\beta)}
    = \frac{\log r}{\beta\lambda} + \frac{r}{\kappa^d(1-r)}
     + \frac{1}{\beta^d}\int_{[0, \infty)^d}a_1^{(\kappa)}(p; r)\,dp.
\]
We also introduce the function $\tilde h_{\kappa}$ on $( 0, \tilde g_0^{(\kappa)-1})$
by
\[
   \tilde h_{\kappa}(r) = \frac{\log r}{\beta\lambda} + \frac{1}{\kappa^d}
         \Tr\frac{r\tilde G_{\kappa}(\beta)}{1-r\tilde G_{\kappa}(\beta)}.
\]
Since $ G_{\kappa}(\beta) \geqslant \tilde G_{\kappa}(\beta)$, $ h_{\kappa} \geqslant \tilde h_{\kappa}$
follows.
Obviously $ h_{\kappa} $ and $ \tilde h_{\kappa}$ are strictly increasing
continuous functions.
Then we have $ r_{\kappa} \leqslant \tilde r_{\kappa}$, because
$ r_{\kappa} $ and $ \tilde r_{\kappa}$ are solutions of $ h_{\kappa}(r) = \mu/\lambda $
and $ \tilde h_{\kappa}(r) = \mu/\lambda $, respectively.
(Recall (\ref{r}) and (\ref{rt}).)

(ii) From $ \tilde h_{\kappa}(\tilde r_{\kappa}) = h_{\kappa}( r_{\kappa}) $,
we have
\[
  \frac{1}{\beta\lambda}\log \frac{\tilde r_{\kappa}}{r_{\kappa}} \leqslant
     \tilde h_{\kappa}(\tilde r_{\kappa}) -  \tilde h_{\kappa}( r_{\kappa}) =
      h_{\kappa}( r_{\kappa}) -  \tilde h_{\kappa}( r_{\kappa})
\]
\[
  = \frac{1}{\kappa^d}\Tr\Big[\frac{1}{ 1-  r_{\kappa}\tilde G_{\kappa}(\beta)}
(G_{\kappa}(\beta) - \tilde G_{\kappa}(\beta)) \frac{r_{\kappa}}{1-r_{\kappa}G_{\kappa}(\beta)}\Big] =O(\kappa^{-d}).
\]
Here we have used Lemma \ref{gt} and the fact that $r_{\kappa}$ is bounded away from 1.
The desired estimate follows.

(iii) Since we already know that $ r_{\kappa}, \tilde r_{\kappa} \to r_* \in (0,1)$,
we get these estimates readily from (\ref{r}) and (\ref{rt}).

(iv) This can be derived by the telescoping together with (ii) and Lemma \ref{gt}.
\hfill$\Box$

\smallskip

Now, let us consider the limit of the ratio
$\tilde\Xi_{\kappa}(\beta, \mu)/\Xi_{\kappa}(\beta, \mu)$
to derive (\ref{norm}). From Lemma \ref{Xi1} and Lemma \ref{tilde_Xi1}, we have
\[
    E_{\beta, r_*}\big[e^{-\langle f, \xi \rangle}\big] = \lim_{\kappa\to\infty}
    \frac{\tilde\Xi_{\kappa}(\beta,\mu)}{\Xi_{\kappa}(\beta,\mu)}
    = \lim_{\kappa\to\infty}\sqrt{\frac{1 + \beta\lambda\kappa^{-d}
     \Tr[r_{\kappa}G_{\kappa}(\beta)(1-r_{\kappa}G_{\kappa}(\beta))^{-2}]}
           {1 + \beta\lambda\kappa^{-d}
     \Tr[\tilde r_{\kappa}\tilde G_{\kappa}(\beta)(1-\tilde r_{\kappa}\tilde G_{\kappa}(\beta))^{-2}]}}
\]
\[
   \times \frac{\Det[1-\tilde r_{\kappa} G_{\kappa}(\beta)]}
    {\Det[1-\tilde r_{\kappa}\tilde G_{\kappa}(\beta)]}
         \frac{\Det[1- r_{\kappa} G_{\kappa}(\beta)]}
              {\Det[1-\tilde r_{\kappa} G_{\kappa}(\beta)]}
        e^{\kappa^d(2\beta\lambda)^{-1}[(\beta\mu - \log\tilde r_{\kappa})^2
                 - (\beta\mu - \log r_{\kappa})^2]}.
\]
Lemma \ref{rtilde_r} yields that the first factor is equal to $1+O(\kappa^{-d})$.
For the second factor, we note that
\[
   \bigg\|\sqrt{1-e^{-f}}\frac{\tilde r_{\kappa} G_{\kappa}(\beta)}
    {1-\tilde r_{\kappa} G_{\kappa}(\beta)}\sqrt{1-e^{-f}}
   - \sqrt{1-e^{-f}}\frac{ r_* G_{\kappa}(\beta)}
    {1- r_* G_{\kappa}(\beta)}\sqrt{1-e^{-f}}\bigg\|_1
\]
\[
    =\bigg\|\sqrt{1-e^{-f}}\sqrt{\frac{ r_* G_{\kappa}(\beta)}
    {1- r_* G_{\kappa}(\beta)}}
    \frac{r_* - \tilde r_{\kappa}}{r_*}
      \frac{1}{1-\tilde r_{\kappa} G_{\kappa}(\beta)}
   \sqrt{\frac{ r_* G_{\kappa}(\beta)}
    {1- r_* G_{\kappa}(\beta)}}\sqrt{1-e^{-f}}\bigg\|_1
\]
\[
       \leqslant \frac{r_* - \tilde r_{\kappa}}{r_*(1-\tilde r_{\kappa})}
       \bigg\|\sqrt{1-e^{-f}}\frac{ r_* G_{\kappa}(\beta)}
    {1- r_* G_{\kappa}(\beta)}\sqrt{1-e^{-f}}\bigg\|_1 \to 0
\]
holds, where $\|\cdot\|_1$ stands for the trace norm.
Together with (\ref{rG}),
\[
    \frac{\Det[1-\tilde r_{\kappa} \tilde G_{\kappa}(\beta)]}
    {\Det[1-\tilde r_{\kappa}G_{\kappa}(\beta)]}
   = \Det[1+\tilde r_{\kappa}( G_{\kappa}(\beta)-\tilde G_{\kappa}(\beta))
    (1-\tilde r_{\kappa} G_{\kappa}(\beta))^{-1}]
\]
\[
        = \Det[1+\sqrt{1-e^{-f}}\frac{\tilde r_{\kappa} G_{\kappa}(\beta)}
    {1-\tilde r_{\kappa} G_{\kappa}(\beta)}\sqrt{1-e^{-f}}] \to
          \Det[1+\sqrt{1-e^{-f}}\frac{ r_{*} G(\beta)}
    {1- r_{*} G(\beta)}\sqrt{1-e^{-f}}]
\]
follows.
Similarly for the third factor, we have
\[
    \frac{\Det[1-\tilde r_{\kappa} G_{\kappa}(\beta)]}
    {\Det[1- r_{\kappa} G_{\kappa}(\beta)]}
        =  \Det\Big[1-\frac{\tilde r_{\kappa}-r_{\kappa}}{r_{\kappa}}
      \frac{r_{\kappa} G_{\kappa}(\beta)}{1- r_{\kappa} G_{\kappa}(\beta)}\Big]
\]
\[
    =\exp\Big( - \frac{\tilde r_{\kappa}-r_{\kappa}}{r_{\kappa}}
    \Tr\Big[\frac{r_{\kappa}G_{\kappa}(\beta)}{1- r_{\kappa} G_{\kappa}(\beta)}\Big] +O(\kappa^{-d})\Big)
    = \exp\Big( - \frac{\tilde r_{\kappa}-r_{\kappa}}{r_{\kappa}}
               \frac{\kappa^d}{\beta\lambda} (\beta\mu -\log r_{\kappa}) +O(\kappa^{-d})
        \Big),
\]
where we have used Lemma \ref{rtilde_r} and (\ref{r}).
It also follows from Lemma \ref{rtilde_r} that the fourth factor is equal to
\[
       \exp\Big[\frac{\kappa^d}{2\beta\lambda}
      (2\beta\mu - \log r_{\kappa}\tilde r_{\kappa})
     \log\Big(1 +\frac{r_{\kappa}-\tilde r_{\kappa}}{\tilde r_{\kappa}}\Big)
      \Big]
     = \exp\Big[ - \frac{\tilde r_{\kappa}-r_{\kappa}}{r_{\kappa}}
               \frac{\kappa^d}{\beta\lambda} (\beta\mu -\log r_{\kappa}) +
      O(\kappa^{-d})  \Big].
\]

Thus we get (\ref{norm}).
Since convergence of generating functionals yields the weak convergence of random point fields,
Theorem \ref{thmA}(i) follows.    \hfill$\Box$
\subsection{ The case {$\mu >\mu_c(\beta)$} (condensed phase). }

Let us put:
\[
    p_j^{(\kappa)}=\frac{r_{\kappa}g_j^{(\kappa)}}{1-r_{\kappa}g_j^{(\kappa)}},
  \quad \varepsilon_j^{(\kappa)}=\log\Big( 1+ \frac{1}{p_j^{(\kappa)}}\Big)
  \quad \mbox{and } \; \delta^{(\kappa)}=\log\Big( 1+ \frac{1}{2p_1^{(\kappa)}}\Big)
\]
for $ j= 0, 1, \cdots.$
Then it follows from (\ref{rs}) and Proposition \ref{cases}(b) that
\[
   s_{\kappa} = \sum_{j=0}^{\infty}p_j^{(\kappa)}, \quad
    1+p_0^{(\kappa)} = \frac{1}{1-r_{\kappa}} =
     \frac{\kappa^d (\beta^d\mu-\zeta(d)\lambda)}{\beta^d\lambda}(1+o(1)) =\hat O(\kappa^d),
     \quad
        \varepsilon_0^{(\kappa)} = \hat O(\kappa^{-d}).
\]
Since $g_1^{(\kappa)} = e^{-\beta/\kappa}$, we have
\[
       p_1^{(\kappa)} = \hat O(\kappa), \quad
        \delta^{(\kappa)}  =  \hat O(\kappa^{-1}),\quad
         \varepsilon_1^{(\kappa)} =  \hat O(\kappa^{-1})
\]
and
\[
     \varepsilon_0^{(\kappa)}< \delta^{(\kappa)} < \varepsilon_1^{(\kappa)}
      \leqslant \varepsilon_2^{(\kappa)} \leqslant \cdots.
\]
Note also that $ s_{\kappa} = \hat O(\kappa^d) $ holds because of (\ref{rs}), (\ref{r}).
By Remark \ref{REMa} and $r_{\kappa}g_j^{\kappa}\in (0,1)$
\begin{equation}
            \sum_{j=1}^{\infty}p_j^{(\kappa)\,2} = O(\kappa^d).
\label{p^2}
\end{equation}
and (\ref{log1}), we get
\[
     \prod_{j=1}^{\infty}\Big( 1-\frac{p_j^{(\kappa)}}{p_0^{(\kappa)}}\Big)
    = \exp\Big(\sum_{j=1}^{\infty}\log\Big( 1-\frac{p_j^{(\kappa)}}{p_0^{(\kappa)}}
         \Big)\Big)
          =\exp\Big(-\frac{s_{\kappa}- p_0^{(\kappa)}}{p_0^{(\kappa)}}
        + O(\kappa^{-d})\Big).
\]
Similarly, we have
\[
    \prod_{j=1}^{\infty}( 1-p_j^{(\kappa)}(e^{\delta^{(\kappa)} }-1))
    = \exp\big(-\sum_{j=1}^{\infty}p_j^{(\kappa)}(e^{\delta^{(\kappa)} }-1)
        + O(\kappa^{d-2})\Big)
\]
\[
     = \exp\big(-\delta^{(\kappa)}(s_{\kappa}-p_0^{(\kappa)}) + O(\kappa^{d-2})\big)
\]
since $p_j^{(\kappa)}(e^{\delta^{(\kappa)} }-1) \leqslant p_1^{(\kappa)}
     (e^{\delta^{(\kappa)} }-1)
       = 1/2 \, $ for $ \, j=1, 2, \cdots$.
\begin{lem}\label{Xi2} For large $\kappa$ we have the asymptotics:
\begin{equation}\label{Asympt-large-mu}
    \Xi_{\kappa}(\beta,\mu)= \sqrt{\frac{2\pi\beta\lambda}{e^2\kappa^d}}
    \frac{\beta^{d-1}e^{\kappa^d(\beta\mu-\log r_{\kappa})^2/2\beta\lambda}}
    {(\beta^d\mu-\zeta(d)\lambda)\Det(1-r_{\kappa}G_{\kappa}(\beta))}(1+o(1)).
\end{equation}
\end{lem}
{\sl Proof : } As in the proof of Lemma \ref{Xi1}, we start with the
integral
\[
   {\cal I} = \int_{\R}dx\,\frac{e^{-is_{\kappa}x-\kappa^dx^2/2\beta\lambda}}
  {\Det[1- (e^{ix}-1)r_{\kappa}G_{\kappa}(\beta)(1-r_{\kappa}G_{\kappa}(\beta))^{-1}]}
\]
\[
  =   \int_{-\infty}^{\infty}dz \, \frac{e^{-is_{\kappa}z-\kappa^dz^2/2\beta\lambda}}
    {\prod_{j=0}^{\infty}(1-(e^{iz}-1)p_j^{(\kappa)})}
  = \bigg(\int_{-\infty}^{\infty} -
      \int_{-\infty-i\delta^{(\kappa)}}^{\infty-i\delta^{(\kappa)}}\bigg)
     + \int_{-\infty-i\delta^{(\kappa)}}^{\infty-i\delta^{(\kappa)}} =
    {\cal I}_1 +{\cal I}_2.
\]
Since   $ 1+ p_0^{(\kappa)}-e^{iz}p_0^{(\kappa)}
           = (1+p_0^{(\kappa)})\big(1-e^{i(z-2n\pi + i\epsilon_0^{(\kappa)})}\big),$
the integrand of ${\cal I}_1$ has simple poles at
$ z= 2n\pi - i\epsilon_0^{(\kappa)} \; ( n\in\Z ).$
Let us calculate ${\cal I}_1$ by means of residues:
\[
    {\cal I}_1 = -2\pi i \sum_{n=-\infty}^{\infty}\mbox{\,Res\,}\Big[
     \frac{1}{(1+p_0^{(\kappa)})\big(1-e^{i(z-2n\pi + i\epsilon_0^{(\kappa)})}\big)}
     \frac{e^{-is_{\kappa}z-\kappa^dz^2/2\beta\lambda}}
    {\prod_{j=1}^{\infty}(1-(e^{iz}-1)p_j^{(\kappa)})} ;
    2n\pi -i\epsilon_0^{(\kappa)}\Big]
\]
\[
    = \frac{2\pi e^{-s_{\kappa}\epsilon_0^{(\kappa)}
       +\kappa^d\epsilon_0^{(\kappa)\,2}/2\beta\lambda}}
       {(1+p_0^{(\kappa)})\prod_{j=1}^{\infty}
      ( 1-p_j^{(\kappa)}/p_0^{(\kappa)})}
      \big( 1 + O\big(e^{-2\pi^2\kappa^d/\beta\lambda}\big)\big)
\]
\[
  = \frac{2\pi\beta^d\lambda e^{-1+O(\kappa^{-d})}}{\kappa^d(\beta^d\mu-\zeta(d)\lambda)}.
\]
Here the pole $z=-i\epsilon_0^{(\kappa)}$ gives the dominant contribution in the second equality.
In the third equality, we have used the relations above this Lemma.

On the other hand, we have
\[
  |{\cal I}_2| \leqslant  \int_{-\infty}^{\infty}dx \,
   \frac{e^{-s_{\kappa}\delta^{(\kappa)}-\kappa^d(x^2-\delta^{(\kappa)\,2})/2\beta\lambda}}
    {\prod_{j=0}^{\infty}|(1-(e^{ix+\delta^{(\kappa)}}-1)p_j^{(\kappa)})|}
\]
\[
    \leqslant \sqrt{\frac{2\pi\beta\lambda}{\kappa^d}}
     \frac{e^{-\hat O(\kappa^{d-1})}}{(e^{\delta^{(\kappa)}}-1)p_0^{(\kappa)} -1}
    =o({\cal I}_1),
\]
where we have used
\[
    \prod_{j=0}^{\infty}|(1-(e^{ix+\delta^{(\kappa)}}-1)p_j^{(\kappa)})|
   \geqslant ((e^{\delta^{(\kappa)}}-1)p_0^{(\kappa)} -1)
        \prod_{j=1}^{\infty}|(1-(e^{\delta^{(\kappa)}}-1)p_j^{(\kappa)})|
\]
\[
   \geqslant ((e^{\delta^{(\kappa)}}-1)p_0^{(\kappa)} -1)
        \exp(-\delta^{(\kappa)}(s_{\kappa}-p_0^{(\kappa)})
      + O(\kappa^{d-2})),
\]
which follows from
\[
      1-(e^{ix+\delta^{(\kappa)}}-1)p^{(\kappa)}_j = (1+ p^{(\kappa)}_j)
        (1-e^{ix+\delta^{(\kappa)}-\varepsilon_j^{(\kappa)}})
\]
and
        $\delta^{(\kappa)}-\varepsilon_0^{(\kappa)} >0, \;
        \delta^{(\kappa)}-\varepsilon_j^{(\kappa)} <0 \; (j=1,2, \cdots)$.
Note also that $\delta^{(\kappa)}p_0^{(\kappa)}= \hat O(\kappa^{d-1})$ holds. From (\ref{pf2}) and
the first equality in (\ref{rs}) one gets desired expression for the asymptotics of $\Xi_{\kappa}(\beta,\mu)$.
\hfill $\Box$

In order to obtain the corresponding asymptotics for $\tilde\Xi_{\kappa}(\beta,\mu)$,
we use the following estimates about
$    \tilde p_j^{(\kappa)}={\tilde r_{\kappa}\tilde g_j^{(\kappa)}}/
{(1-\tilde r_{\kappa}\tilde g_j^{(\kappa)})} \quad ( j = 0, 1, \cdots ) $.

\begin{lem}
\[
    1-\tilde r_{\kappa}\tilde g_0^{(\kappa)} = \frac{\beta^d\lambda(1+o(1))}
            {\kappa^d(\beta^d\mu - \zeta(d)\lambda)},
         \quad |1-\tilde r_{\kappa}| = O(\kappa^{-d/2}),
\]
\[
       \tilde p_0^{(\kappa)} = \hat O(\kappa^d), \quad
       \tilde p_1^{(\kappa)} =  \hat O(\kappa), \quad
       \sum_{j=1}^{\infty}\tilde p_j^{(\kappa)} = \hat O(\kappa^d) ,\quad
      \sum_{j=1}^{\infty}\tilde p_j^{(\kappa)\,2} = O(\kappa^d).
\]
\label{tilde_p}
\end{lem}
{\sl Proof : } Proposition \ref{cases}(b) , Lemma \ref{rtilde_r}(i) and
Lemma \ref{gt} yield
\begin{equation}
    1 - \hat O(\kappa^{-d})= r_{\kappa} \leqslant \tilde r_{\kappa} <
        \tilde g_0^{(\kappa)\,-1} = 1 + \hat O(\kappa^{-d/2}),
\label{rr}
\end{equation}
which implies $ |1- \tilde r_{\kappa}| = O(\kappa^{-d/2}) $.
Note that the argument which shows $ r_{\kappa} \leqslant  \tilde r_{\kappa} $
in the proof of Lemma \ref{rtilde_r} is also valid for the present case.
In the variational formula
\[
  \tilde g_1^{(\kappa)} = \sup_{\psi\perp \tilde \Omega}\frac{(\psi, \tilde G_{\kappa}(\beta)\psi)}
     {(\psi, \psi)}
        = \sup_{\psi\perp \tilde \Omega}\frac{[(\psi, G_{\kappa}(\beta)\psi)
    - (\psi, D_{\kappa}\psi)]}{(\psi, \psi)},
\]
we can use as $\psi$ a linear combination of two eigenfunctions
      $ \kappa^{-d/4}\phi_s(x/\sqrt{\kappa})$  of $ G_{\kappa}(\beta)$
perpendicular to $\tilde \Omega$ (e.g., with $s = (1,0,0, \cdots)$ and $s=( 0,1,0, \cdots)$).
Then we get $ \tilde g_1^{\kappa} \geqslant 1- \hat O(\kappa^{-1})$.
Together with
$ \tilde g_1^{(\kappa)} \leqslant g_1^{(\kappa)} = 1-\hat O(\kappa^{-1}) $,
$ \tilde g_1^{(\kappa)} = 1-\hat O(\kappa^{-1}) $ follows.
Thus $ \hat O(\kappa) = \tilde p_1^{(\kappa)} \geqslant \tilde p_2^{(\kappa)} \geqslant \cdots $
holds.

Now we get
\[
    |\sum_{j=1}^{\infty}\tilde p_j^{(\kappa)}
   - \sum_{j=1}^{\infty}\tilde g_j^{(\kappa)}/( 1- \tilde g_j^{(\kappa)})|
   =  |1-\tilde r_{\kappa}| \sum_{j=1}^{\infty}\tilde g_j^{(\kappa)}
    /\big(( 1- \tilde r_{\kappa}\tilde g_j^{(\kappa)})( 1- \tilde g_j^{(\kappa)})\big)
\]
\[
   =O(\kappa^{1-d/2})\sum_{j=1}^{\infty}\tilde g_j^{(\kappa)}/( 1- \tilde g_j^{(\kappa)}),
\]
which implies
\[
     \sum_{j=1}^{\infty}\tilde p_j^{(\kappa)}
   = (1+ O(\kappa^{1-d/2}))\sum_{j=1}^{\infty}\tilde g_j^{(\kappa)}/
   ( 1- \tilde g_j^{(\kappa)}).
\]
On the other hand, because
\[
     |\sum_{j=1}^{\infty}\frac{\tilde g_j^{(\kappa)}}{ 1- \tilde g_j^{(\kappa)}}
          - \sum_{j=1}^{\infty} \frac{g_j^{(\kappa)}}{ 1-  g_j^{(\kappa)}}|
   \leqslant \sum_{j=1}^{\infty}\frac{ g_j^{(\kappa)}- \tilde g_j^{(\kappa)}}
    {(1-  g_1^{(\kappa)})( 1- \tilde g_1^{(\kappa)})}
    \leqslant \frac{\Tr D_{\kappa}}{(1-  g_1^{(\kappa)})( 1- \tilde g_1^{(\kappa)})}= O(\kappa^2),
\]
we have
\[
     \frac{1}{\kappa^d}\sum_{j=1}^{\infty}\frac{\tilde g_j^{(\kappa)}}
      { 1- \tilde g_j^{(\kappa)}} =
      \frac{1}{\kappa^d}\sum_{j=1}^{\infty}\frac{ g_j^{(\kappa)}}
      { 1-  g_j^{(\kappa)}} +O(\kappa^{2-d}) = \frac{\zeta(d)}{\beta^{d}} + o(1),
\]
where we recall Remark \ref{REMa}.
Thus we have $ \kappa^{-d}\sum_{j=1}^{\infty}\tilde p_j^{(\kappa)}
     = \beta^{-d}\zeta(d) + o(1).$
Using (\ref{rt}), we get
\[
     \frac{\tilde r_{\kappa}\tilde g_0^{(\kappa)}}
    {\kappa^d(1-\tilde r_{\kappa}\tilde g_0^{(\kappa)})}
     = -\frac{\log \tilde r_{\kappa}}{\beta\lambda} + \frac{\mu}{\lambda}
     -\frac{1}{\kappa^d} \sum_{j=1}^{\infty}  \tilde p_j^{(\kappa)}
      = \frac{\mu}{\lambda} -\frac{\zeta(d)}{\beta^d} +o(1)
\]
which yields the first and the third equality.

To prove the remaining last bound, it is enough to show that
\begin{equation}
  \tilde p_j^{(\kappa)}\leqslant 2 p_j^{(\kappa)} \quad ( j= 1, 2, \cdots )
\label{p2p}
\end{equation}
hold for large enough $\kappa$, because of
$ \sum_{j=1}^{\infty}  p_j^{(\kappa)\, 2} = O(\kappa^d)$.
In fact, in the expression
\[
   \tilde p_j^{(\kappa)} = \frac{ r_{\kappa} \tilde g_j^{(\kappa)}}
   {1- r_{\kappa} \tilde g_j^{(\kappa)}}
  \frac{1+(\tilde r_{\kappa}-r_{\kappa})/r_{\kappa}}
     {1-(\tilde r_{\kappa}-r_{\kappa})\tilde g_j^{(\kappa)}/
       (1-r_{\kappa}\tilde g_j^{(\kappa)})},
\]
 $ (\tilde r_{\kappa}-r_{\kappa})/r_{\kappa} = O(\kappa^{-d/2}) \, $ and
            $ \, |(\tilde r_{\kappa}-r_{\kappa}) \tilde g_j^{(\kappa)}/
             (1-r_{\kappa}\tilde g_j^{(\kappa)})| \leqslant
    (\tilde r_{\kappa}-r_{\kappa})/
             (1-r_{\kappa} g_1^{(\kappa)}) = O(\kappa^{1-d/2}) $
hold.
Because of $ \tilde g_j^{(\kappa)} \leqslant g_j^{(\kappa)}$, we also have
$ r_{\kappa}\tilde g_j^{(\kappa)}/(1-r_{\kappa}\tilde g_j^{(\kappa)}) \leqslant
    p_j^{(\kappa)} $.
Thus we get (\ref{p2p}).  \hfill$\Box$

It is obvious now that the next Lemma can be derived along the same line of reasoning
as the proof of Lemma \ref{Xi2}.
\begin{lem}\label{tilde_Xi2}
For large $\kappa$ one gets the asymptotics:
\begin{equation}\label{Asympt-tilda}
   \tilde \Xi_{\kappa}(\beta,\mu)= \sqrt{\frac{2\pi\beta\mu}{e^2 \kappa^d}}
    \frac{\beta^{d-1}e^{\kappa^d(\beta\mu-\log\tilde  r_{\kappa})^2/2\beta\lambda}}
    {(\beta^d\mu-\zeta(d)\lambda)\Det(1-\tilde r_{\kappa}\tilde G_{\kappa}(\beta))}(1+o(1)).
\end{equation}
\end{lem}

\medskip

In order to calculate the limit of
   $\tilde\Xi_{\kappa}(\beta,\mu)/\Xi_{\kappa}(\beta,\mu)$,
we use the following lemma, where we put
\[
      \hat{g}_0 ^{(\kappa)} := (\Omega_{\kappa}, \tilde G_{\kappa}(\beta) \Omega_{\kappa}) +
          \tilde r_{\kappa} (\Omega_{\kappa}, \tilde G_{\kappa}(\beta)Q_{\kappa}
       (1-\tilde r_{\kappa} Q_{\kappa}\tilde G_{\kappa}(\beta)Q_{\kappa})^{-1}
         Q_{\kappa}\tilde G_{\kappa}(\beta)\Omega_{\kappa}).
\]
\begin{lem}\label{last}
For large $\kappa$ one gets:
\begin{eqnarray*}
{\rm(i)}& \qquad \; \tilde r_{\kappa}- r_{\kappa}=(1-\tilde g_0^{(\kappa)})(1+o(1))
         = \hat O(\kappa^{-d/2}),
\\
{\rm (ii)}& \qquad 1-\tilde r_{\kappa}\hat{g}_0^{(\kappa)} =
         (1-\tilde r_{\kappa}\tilde g_0^{(\kappa)})(1+o(1)).
\end{eqnarray*}
\end{lem}
{\sl Proof:} From Lemma \ref{tilde_p} and Proposition \ref{cases}(b), we have
   $ \tilde r_{\kappa}\tilde g_0^{(\kappa)} -r_{\kappa} = o(\kappa^{-d})$.
Hence (i) follows from
$ \tilde r_{\kappa} -r_{\kappa} = \tilde r_{\kappa}(1-\tilde g_0^{(\kappa)})
  + \tilde r_{\kappa}\tilde g_0^{(\kappa)} -r_{\kappa},
      \tilde r_{\kappa} = 1+O(\kappa^{d/2})$ and Lemma \ref{gt}.

By virtue of Lemma \ref{gt} we get
\[
     \tilde g_0^{(\kappa)}- \hat{g}_0^{(\kappa)} =
    (W_{\kappa}^*\Omega_{\kappa}, W_{\kappa}^*Q_{\kappa}
         [(\tilde g_0^{(\kappa)}- Q_{\kappa}\tilde G_{\kappa}(\beta)Q_{\kappa})^{-1}
        - (\tilde r_{\kappa}^{-1} - Q_{\kappa}\tilde G_{\kappa}(\beta)Q_{\kappa})^{-1}]
          Q_{\kappa}W_{\kappa}W_{\kappa}^*\Omega_{\kappa})
\]
\begin{equation}
   =  (W_{\kappa}^*\Omega_{\kappa}, W_{\kappa}^*Q_{\kappa}
         (\tilde g_0^{(\kappa)}- Q_{\kappa}\tilde G_{\kappa}(\beta)Q_{\kappa})^{-1}
        (\tilde r_{\kappa}^{-1}-\tilde g_0^{(\kappa)})
        (\tilde r_{\kappa}^{-1} - Q_{\kappa}\tilde G_{\kappa}(\beta)Q_{\kappa})^{-1}
          Q_{\kappa}W_{\kappa}W_{\kappa}^*\Omega_{\kappa}),
\label{g-g'}
\end{equation}
which yields $ \tilde g_0^{(\kappa)} - \hat{g}_0^{(\kappa)} \geqslant 0 $ since
 $ \tilde r_{\kappa} < \tilde g_0^{(\kappa)\,-1} $.
Recall that $||\tilde X || \leqslant 1$ and
$ ||W_{\kappa}^*\Omega_0^{(\kappa)}|| = O(\kappa^{-d/4})$.
( See (\ref{XX'}) and its next line.)
Then we also get from (\ref{g-g'}) that
\[
        \tilde g_0^{(\kappa)}- \hat{g}_0^{(\kappa)} \leqslant
       \frac{1-\tilde r_{\kappa}\tilde g_0^{(\kappa)}}
       {1-\tilde r_{\kappa}g_1^{(\kappa)}}
       (W_{\kappa}^*\Omega_0^{(\kappa)}, W_{\kappa}^*Q_{\kappa}
         (\tilde g_0^{(\kappa)}- Q_{\kappa}\tilde G_{\kappa}(\beta)Q_{\kappa})^{-1}
          Q_{\kappa}W_{\kappa}W_{\kappa}^*\Omega_0^{(\kappa)})
\]
\[
    \leqslant \frac{1-\tilde r_{\kappa}\tilde g_0^{(\kappa)}}
       {1-\tilde r_{\kappa}g_1^{(\kappa)}}||\tilde X ||
      ||W_{\kappa}^*\Omega_0^{(\kappa)}||^2
        = (1-\tilde r_{\kappa}\tilde g_0^{(\kappa)})O(\kappa^{1-d/2}).
\]
Hence we obtain the asymptotics (ii):
\[  \hskip15mm
    1-\tilde r_{\kappa} \hat{g}_0^{(\kappa)} =
    1-\tilde r_{\kappa}\tilde g_0^{(\kappa)} + \tilde r_{\kappa}
      (\tilde g_0^{(\kappa)} - \hat{g}_0^{(\kappa)})
    = (1-\tilde r_{\kappa}\tilde g_0^{(\kappa)})(1+o(1)).
  \hskip2cm \Box
\]

Now, taking into account (\ref{Asympt-large-mu}) and (\ref{Asympt-tilda}), we can find the asymptotics
of the generating functional (\ref{EK1}):
\[
    E_{\kappa, \beta, \mu}\big[e^{-\langle f, \xi \rangle}\big] =
   \frac{\tilde\Xi_{\kappa}(\beta, \mu)}{\Xi_{\kappa}(\beta, \mu)}
    = \exp\Big(\frac{\kappa^d}{2\beta\lambda}(2\beta\mu - \log r_{\kappa}
        \tilde r_{\kappa}) \log \frac{r_{\kappa}}{\tilde r_{\kappa}}\Big)
       \frac{\Det[1-\tilde r_{\kappa} Q_{\kappa}\tilde G_{\kappa}(\beta)Q_{\kappa}]}
    {\Det[1-\tilde r_{\kappa}\tilde G_{\kappa}(\beta)]}
\]
\begin{equation}
   \times  \frac{\Det[1-\tilde r_{\kappa} Q_{\kappa}G_{\kappa}(\beta)Q_{\kappa}]}
             {\Det[1-\tilde r_{\kappa} Q_{\kappa}\tilde G_{\kappa}(\beta)Q_{\kappa}]}
         \, \frac{\Det[1- r_{\kappa} Q_{\kappa}G_{\kappa}(\beta)Q_{\kappa}]}
              {\Det[1-\tilde r_{\kappa} Q_{\kappa}G_{\kappa}(\beta)Q_{\kappa}]}
         \, \frac{\Det[1- r_{\kappa} G_{\kappa}(\beta)]}
         {\Det[1- r_{\kappa} Q_{\kappa}G_{\kappa}(\beta)Q_{\kappa}]}(1+o(1)).
\label{5factors}
\end{equation}

By virtue of Lemma \ref{last}(i), for the exponent of the first factor, we have
\[
       \frac{\kappa^d}{2\beta\lambda}(2\beta\mu - \log r_{\kappa}\tilde r_{\kappa})
                 \log \frac{r_{\kappa}}{\tilde r_{\kappa}}
            = -\frac{\mu\kappa^d}{\lambda}(1-\tilde g_0^{(\kappa)}) (1+o(1)) \, .
\]

For the second factor, we use the Feshbach formula, which claims
\[
  \Det A= \Det B\, \Det(C-K^TB^{-1}K),
\]
where
\[
    A =
\begin{pmatrix}
             B  & -K     \\  -K^T &  C
\end{pmatrix}
             =
\begin{pmatrix}
           1  &  0     \\  -K^TB^{-1} & 1
\end{pmatrix}
\begin{pmatrix}
           B  &  0     \\   0  & C-K^TB^{-1}K
\end{pmatrix}
\begin{pmatrix}
         1  & -B^{-1}K \\ 0  & 1
\end{pmatrix}.
\]
This formula and Lemma \ref{last}(ii) yield
\begin{eqnarray*}
&& \frac{\Det[1-\tilde r_{\kappa} Q_{\kappa}\tilde G_{\kappa}(\beta)Q_{\kappa}]}
{\Det[1-\tilde r_{\kappa}\tilde G_{\kappa}(\beta)]} = \\
&&\frac{1}{1-  \tilde r_{\kappa} (\Omega_0^{(\kappa)},
\tilde G_{\kappa}(\beta) \Omega_0^{(\kappa)})
-  (\Omega_0^{(\kappa)}, \tilde r_{\kappa}\tilde G_{\kappa}(\beta)Q_{\kappa}
(1-\tilde r_{\kappa} Q_{\kappa} \tilde G_{\kappa}(\beta)Q_{\kappa})^{-1}
Q_{\kappa}\tilde r_{\kappa} \tilde G_{\kappa}(\beta)\Omega_0^{(\kappa)})}  \\
&& = 1/(1- \tilde r_{\kappa}\hat{g}_0^{(\kappa)}) =
(1+o(1))/( 1- \tilde r_{\kappa}\tilde g_0^{(\kappa)}).
\end{eqnarray*}
Since
\[
       \frac{\Det[1- r_{\kappa} G_{\kappa}(\beta)]}
         {\Det[1- r_{\kappa} Q_{\kappa}G_{\kappa}(\beta)Q_{\kappa}]} = 1-r_{\kappa} ,
\]
then Proposition \ref{cases}(b) and Lemma \ref{tilde_p} yields for the product of factors in (\ref{5factors}):
\[
   \mbox{(the 2nd factor)} \times \mbox{(the last factor)} \to 1
\]
in the limit $\kappa \to \infty$.

Now, since Lemma \ref{tilde_p} and (\ref{QG}) give
\[
  || \sqrt{1-e^{-f}}\frac{\tilde r_{\kappa}Q_{\kappa}G_{\kappa}(\beta)Q_{\kappa}}
        {1-\tilde r_{\kappa} Q_{\kappa}G_{\kappa}(\beta)Q_{\kappa}}\sqrt{1-e^{-f}}
    -K_f ||_1
\]
\[
    \leqslant || \sqrt{1-e^{-f}}\frac{\tilde r_{\kappa}Q_{\kappa}G_{\kappa}(\beta)Q_{\kappa}}
        {1-\tilde r_{\kappa} Q_{\kappa}G_{\kappa}(\beta)Q_{\kappa}}\sqrt{1-e^{-f}}
    -\sqrt{1-e^{-f}}\frac{Q_{\kappa}G_{\kappa}(\beta)Q_{\kappa}}
    {1- Q_{\kappa}G_{\kappa}(\beta)Q_{\kappa}}\sqrt{1-e^{-f}}\|_1
\]
\[
     + ||\sqrt{1-e^{-f}}\frac{Q_{\kappa}G_{\kappa}(\beta)Q_{\kappa}}
    {1- Q_{\kappa}G_{\kappa}(\beta)Q_{\kappa}}\sqrt{1-e^{-f}}-K_f ||_1
\]
\[
    \leqslant | \tilde r_{\kappa} -1|\, ||\sqrt{1-e^{-f}}\frac{Q_{\kappa}G_{\kappa}(\beta)Q_{\kappa}}
    {1- Q_{\kappa}G_{\kappa}(\beta)Q_{\kappa}}\sqrt{1-e^{-f}}||_1
      ||(1-\tilde r_{\kappa}Q_{\kappa}G_{\kappa}(\beta)Q_{\kappa})^{-1}||
\]
\[
    + ||\sqrt{1-e^{-f}}\frac{Q_{\kappa}G_{\kappa}(\beta)Q_{\kappa}}
    {1- Q_{\kappa}G_{\kappa}(\beta)Q_{\kappa}}\sqrt{1-e^{-f}} - K_f||_1
      \to 0
\]
for $ \kappa \to \infty$, we obtain the limit:
\[
    \frac{\Det[1-\tilde r_{\kappa} Q_{\kappa}G_{\kappa}(\beta)Q_{\kappa}]}
             {\Det[1-\tilde r_{\kappa} Q_{\kappa}\tilde G_{\kappa}(\beta)Q_{\kappa}]}
    = \frac{1}{\Det[1+\tilde r_{\kappa} Q_{\kappa}(G_{\kappa}(\beta)-\tilde G_{\kappa}(\beta))
        Q_{\kappa}(1-\tilde r_{\kappa} Q_{\kappa}G_{\kappa}(\beta)Q_{\kappa})^{-1}]}
\]
\[
     =\Det[1+\tilde r_{\kappa} \sqrt{1-e^{-f}}Q_{\kappa}G_{\kappa}(\beta)Q_{\kappa}
         (1-\tilde r_{\kappa} Q_{\kappa}G_{\kappa}(\beta)Q_{\kappa})^{-1}\sqrt{1-e^{-f}}]^{-1}
      \to \Det[1+K_f]^{-1}
\]
for the third factor in (\ref{5factors}).
Here we have used the \textit{cyclicity} of the Fredholm determinant.

Lemma \ref{last}(i), (\ref{p^2}), (\ref{rs}) and Proposition \ref{cases}(b) yield
\[
      \frac{\Det[1- r_{\kappa} Q_{\kappa}G_{\kappa}(\beta)Q_{\kappa}]}
              {\Det[1-\tilde r_{\kappa} Q_{\kappa}G_{\kappa}(\beta)Q_{\kappa}]}
          = \frac{1}{\Det[1- (\tilde r_{\kappa}- r_{\kappa})
         Q_{\kappa}G_{\kappa}(\beta)Q_{\kappa}(1-r_{\kappa} Q_{\kappa}G_{\kappa}(\beta)Q_{\kappa})^{-1}]}
\]
\[
    =   \exp\Big(\frac{\tilde r_{\kappa}- r_{\kappa}}{r_{\kappa}}
       \Tr \frac{ r_{\kappa} Q_{\kappa}G_{\kappa}(\beta)Q_{\kappa}}
        {1-r_{\kappa} Q_{\kappa}G_{\kappa}(\beta)Q_{\kappa}} + O(1)\Big)
    =   \exp\Big((1-\tilde g_0^{(\kappa)})\Big[\kappa^d
         \frac{\beta\mu-\log r_{\kappa}}{\beta\lambda}-\frac{r_{\kappa}}{1-r_{\kappa}}
             \Big] + O(1) \Big)
\]
\[
    =\exp\{(1-\tilde g_0^{(\kappa)})\kappa^d \zeta(d)(1+o(1))/\beta^d \}.
\]
Thus, by Lemma \ref{gt} and (\ref{QG}) we get for the product in (\ref{5factors}):
\[
   \mbox{(the 1st factor)} \times \mbox{(the 4th factor)}
\]
\[
        = \exp\Big( - \frac{\kappa^{d/2}(1+o(1))}{{\pi^{d/2}}}
           \frac{\beta^d\mu - \zeta(d)\lambda}{\beta^d\lambda}
          (\sqrt{1-e^{-f}}, (1+K_f)^{-1}\sqrt{1-e^{-f}})\Big) .
\]

Now Theorem \ref{thmA}(ii) follows by collecting in (\ref{5factors}) the asymptotics of factors
that we find above. \hfill$\Box$
\section{ Proof of Theorem \ref{thmB}}
We start with the grand-canonical expectation value of the total number
of MF interacting bosons in the WHT (\ref{s-a-glob-dens}):
\begin{equation*}
N_{\kappa, \lambda}(\beta, \mu) := \kappa^d \, {\rho}_{\kappa,\lambda}^{(tot)}(\beta,\mu)=
\frac{1}{\beta \, \Xi_{\kappa}(\beta, \mu)}
\frac{\partial}{\partial \mu}\Xi_{\kappa}(\beta, \mu) \ ,
\end{equation*}
where we use for $\Xi_{\kappa}(\beta, \mu) $ the expression (\ref{pf1}) after the
$z$-integration where the values of  $s$ and $r$ are not fixed yet.
The differentiation with respect to $\mu$ can be converted into differentiation
with respect to $x$ in the Fredholm determinant.
Then integrating by parts we obtain
\begin{equation*}
       N_{\kappa, \lambda}(\beta, \mu)= s - \frac{i\kappa^d}{\beta\lambda}R(r,s)  \ ,
\end{equation*}
where
\begin{eqnarray}\label{R}
R(r,s)&:=& \int_{\R}dx\,\frac{x \, e^{-isx-\kappa^dx^2/2\beta\lambda}}
{\Det[1- (e^{ix}-1)rG_{\kappa}(\beta)(1-rG_{\kappa}(\beta))^{-1}]} \\
&\times& \Big[ \int_{\R}dx\,\frac{ e^{-isx-\kappa^dx^2/2\beta\lambda}}
  {\Det[1- (e^{ix}-1)rG_{\kappa}(\beta)(1-rG_{\kappa}(\beta))^{-1}]} \Big]^{-1} \ .\nonumber
\end{eqnarray}
Then we put $s = s_{\kappa}$ and $r=r_{\kappa}$ such that (\ref{rs}) holds, and
if we prove $R(r_{\kappa}, s_{\kappa})=o(1)$, then (\ref{Blim}) follows as
a consequence of this asymptotics.

To this end notice that for the case (i) $\mu < \mu_c(\beta)$, we get
$ R(r_{\kappa}, s_{\kappa}) = O(\kappa^{-d/2}) $
from the estimates in the proof of Lemma \ref{Xi1}.

For the case (ii) $\mu > \mu_c(\beta)$, using the notations of the proof of Lemma \ref{Xi2},
one obtains for the first factor in (\ref{R}):
\begin{equation*}
\int_{\R}dx\,\frac{x \, e^{-is_{\kappa}x-\kappa^dx^2/2\beta\lambda}}
{\Det[1- (e^{ix}-1)r_{\kappa}G_{\kappa}(\beta)(1-r_{\kappa}G_{\kappa}(\beta))^{-1}]}
= -i\varepsilon_0{\cal I}_1 + o({\cal I}_1) \ .
\end{equation*}
Therefore, we get that $R(r_{\kappa}, s_{\kappa})=O(\kappa^{-d})$.

The other properties stated in the Theorem \ref{thmB} follow straightforwardly from Section \ref{2-regimes}
and the line of reasoning developed for the proof of Proposition \ref{cases}.   \hfill$\Box$

\section{ Concluding Remarks and Conjectures}
In the present paper we consider a model of mean-field interacting boson gas in traps
described by the harmonic potential.
For this model we study the position distribution of the constituent bosons in the WHT limit
by means of the RPF method.

It is shown that there are two phases distinguished by the boson condensation.
In one domain of parameters, the resulting generating functional for the RPF is the same
as for the non-interacting boson gas, for unconventional values of the IBG parameters.
Whereas in another domain, the generating functional describes  divergence of the density  due to
the localization of macroscopic number of particles.

Our results are obtained via analysis of the generating functional.
We do not intend to start the analysis using the characteristic functional here.
However, we would like to mention a topic on the central limit theorem as a conjecture.

Let us consider $ E_{\kappa,\beta,\mu}[e^{-\langle f, \xi \rangle}] $ for small $ f\in C_0(\R^d)$
in the sense of the sup-norm $\| \cdot \|_{\infty}$. By Theorem \ref{thmA}(ii) we obtain:
\begin{eqnarray*}
&&E_{\kappa,\beta,\mu}\big[e^{-\langle f, \xi \rangle}\big] =  \\
&& \exp \left\{-\frac{\kappa^{d/2}}{\pi^{d/2}}
\frac{\mu-\mu_{\lambda, c}(\beta)}{\lambda}\left[\int_{\mathbb{R}^d} dx \ f(x)
+ \frac{1}{2} \langle f, (1+G(\beta))(1-G(\beta))^{-1}f \rangle + \cdots \right]\right\} \ .
\end{eqnarray*}
This leads us to conclusion that the following claim might be true:
\begin{conjecture} \label{Con1}
Let $\xi_{\kappa}$ be the random point measure on $\R^d$ with distribution given by
$\nu_{\kappa,\beta,\mu}$ for $\kappa >0$, and $\mathfrak{L}$ be the Lebesgue measure on $\R^d$.
Then for $\kappa \to \infty$ the random field
\begin{equation*}
\frac{\pi^{d/4}}{\kappa^{d/4}}\Big(\frac{\lambda}{\mu-\mu_{\lambda,c}(\beta)}\Big)^{1/2}
\Big( \xi_{\kappa} - \frac{\kappa^{d/2}}{\pi^{d/2}}
\frac{\mu-\mu_{\lambda,c}(\beta)}{\lambda}\ {\mathfrak{L}}\Big)
\end{equation*}
converges in distribution to the Gaussian random field on $\R^d$ with covariance
\begin{equation*}
(1+G(\beta))(1-G(\beta))^{-1} \ .
\end{equation*}
\end{conjecture}

We finish by some remarks about the method of the RPF approach to the BEC in the WHT limit
used in the present paper:

\noindent(i) It could be applied to  a general ``non-quadratic" mean-field interaction
$U_\Phi:= \kappa^{d} \Phi(n/\kappa^{d})$, where $\Phi: x\in\mathbb{R} \mapsto \mathbb{R}$ is a
piece-wise differentiable continuous function bounded from below {\rm{\cite{TZ}}},
as well as to the van der Waals particle interaction, which is more local than
the mean-field {\rm{\cite{deS-Z}}}.
We guess that for this kind of interaction the particle distribution will spread as
$\kappa^{\alpha}$ with some large $\alpha$ even for condensed particles.

\noindent (ii)  The method has to be compared with the scaled external field perturbation of BEC
considered in \cite{deS-Z}, \cite{Pu}. We suppose that this could clarify the concept of the
choice of ``effective" volume, since it is important for description of the local particle
density measured in the WHT limit BEC experiments as well as its definition of the
mean-field interaction \cite{DGPS}, \cite{LSSY}, \cite{PeSm}.

\bigskip

\noindent \textbf{Acknowledgements}

\bigskip

\noindent H.T. thanks  MEXT for the financial support
under the Grant-in-Aid for Scientific Research No. 17654021. V.A.Z. is thankful to
Mathematical Department of the Kanazawa University for a warm hospitality and financial support.

\newpage
\appendix
\section{Miscellaneous formulae }
\begin{eqnarray}
   \frac{x}{1-e^{-x}}\leqslant 1+x \leqslant 2\vee (2x) \qquad  \mbox{for }& \; x > 0.
\label{kei1}\\
   0 \leqslant  \bigg( \frac{2x}{1-e^{-2x}}\bigg)^{d/2} -1 \leqslant Ax
  \quad  \mbox{for } &\; x \in [0,1],
\label{kei2}\\
    0 \leqslant  \bigg( \frac{1}{1-e^{-2x}}\bigg)^{d/2} -1 \leqslant Be^{-2x}
  \quad  \mbox{for } &\; x \geqslant 1,
\label{kei3}
\end{eqnarray}
where  $A$ and $B$ are constants depending only on $d$.
\begin{eqnarray}
    1-e^{-x} \leqslant x \qquad  \mbox{ for }& \; x \geqslant 0.
\label{exp1}\\
 e^{-x} - e^{-y} \leqslant \frac{1}{e}\bigg(\frac{y}{x}-1\bigg) \qquad
\mbox{ for }& \; 0 < x \leqslant y.
\label{exp2}
\end{eqnarray}
\begin{eqnarray}
 \tanh x \leqslant x \wedge 1 \qquad &\mbox{for } \; x \geqslant 0.
\label{th1}\\
| \coth x -1 |\leqslant \frac{2e}{e-1}e^{-x} \qquad &\mbox{for } \; x \geqslant 1.
\label{th2}
\end{eqnarray}
\begin{eqnarray}
  1\leqslant \frac{\sinh x}{x} \leqslant 1 + x^2 \qquad  \mbox{for }& \; x\in[0,1].
\label{sh1}\\
 \frac{1}{\sinh x} \leqslant \frac{2}{1-1/e^2}e^{-x} \qquad  \mbox{for }&
 \; x \geqslant 1.
\label{sh2}\\
  |-\log(1-x) - x| \leqslant \frac{x^2}{2(1-x)} \qquad  \mbox{for }&
 \; x < 1.
\label{log1}
\end{eqnarray}

\newpage

\end{document}